\documentclass{aa}

\usepackage{natbib}
\usepackage{graphicx}	
\graphicspath{{figures/}}	
\usepackage{amsmath}	
\usepackage{amsfonts}
\usepackage{amssymb}
\usepackage{txfonts}
\usepackage{hyperref}
\hypersetup{
  colorlinks=true, 
  urlcolor=blue, 
  linkcolor=blue, 
  citecolor=blue 
}

\usepackage{xcolor}
\usepackage{multirow}

\def\chem{\texttt{chem}}

\def\nh{N_{\rm H}}
\def\dlg{\Delta\log g}
\def\Teff{T_{\rm eff}}

\begin{document}

\title{Soft X-ray emission from the classical nova AT~2018bej}

\author{A.~Tavleev\inst{1} \and L.~Ducci\inst{1,2} \and V.~F.~Suleimanov\inst{1} \and C.~Maitra\inst{3} \and K.~Werner\inst{1} \and A.~Santangelo\inst{1} \and V.~Doroshenko\inst{1}}

\institute{Institut f\"ur Astronomie und Astrophysik, Kepler Center for Astro and Particle Physics, Universit\"at T\"ubingen, Sand 1, 72076 T\"ubingen, Germany \and
ISDC Data Center for Astrophysics, Université de Genève, 16 chemin d’Écogia, 1290, Versoix, Switzerland \and
Max-Planck-Institut f\"ur extraterrestrische Physik, Gie{\ss}enbachstra{\ss}e 1, 85748 Garching, Germany \\
          \email{tavleev@astro.uni-tuebingen.de}
         }

\date{Received xxx / accepted xxx}

  \abstract
  {Classical novae are known to demonstrate a supersoft X-ray source (SSS) state following outbursts. This state is associated with residual thermonuclear burning on the white dwarf~(WD) surface. During its all-sky survey~(eRASS1), the eROSITA telescope onboard the Spectrum-Roentgen-Gamma observatory discovered a bright new SSS, whose position is consistent with the known classical nova AT~2018bej in the Large Magellanic Cloud. There were two eROSITA spectra obtained during eRASS1 and eRASS2 monitoring epochs and one \textit{XMM-Newton} grating spectrum close to the eRASS1 epoch.}
  {We aim to describe the eROSITA and follow-up \textit{XMM-Newton} spectra of AT~2018bej with our local thermodynamic equilibrium~(LTE) atmosphere models. We focused on the evolution of the hot WD properties between the eRASS1 and eRASS2 epochs, especially on the change of the carbon abundance.}
  {A grid of LTE~model atmosphere spectra were calculated for different values of the effective temperature (from $\Teff = 525$ to $700\,\rm kK$ in steps of $25\,\rm kK$), surface gravity (six values) and chemical composition assuming approximately equal hydrogen and helium number fractions and five different values of carbon and nitrogen abundances.}
  {Both eRASS1 and XMM\, $0.3-0.6$~keV spectral analyses yield a temperature of the WD of $\Teff {\sim}\,600\, \rm kK$ and a WD radius of~$8000-8700\,\rm km$. Simultaneous fitting of the eROSITA spectra for two epochs (eRASS1 and eRASS2) with a common WD~mass parameter demonstrates a decrease in $\Teff$
  accompanied by an increase in the WD~radius 
  and a decrease in the carbon abundance. However, these changes are marginal and coincide within errors.
  The derived WD mass is estimated to be $1.05-1.15\, M_\sun$.}
  {We traced a minor evolution of the source on a half-year timescale accompanied by a decrease in carbon abundance and concluded that LTE~model atmospheres can be used to analyse the available X-ray spectra of classical novae during their SSS~stage.
  }
  
   \authorrunning{Tavleev et al.}
   \titlerunning{A supersoft X-ray emission from the classical nova AT~2018bej}

   \maketitle
%

\section{Introduction}

A new X-ray source was discovered by eROSITA during the first part of its all-sky survey, whose position is consistent with the known classical nova (CN) AT~2018bej~in the Large Magellanic Cloud \citep[LMC,][]{telegram_xray, telegram_optic}. The X-ray spectrum was thermal with low temperature of~${\sim}30$~eV and high luminosity~$>10^{37}$\,erg\,s$^{-1}$. These properties led to conclusion that the super-soft source~(SSS) phase of CNe was observed for this nova~\citep{telegram_xray}.

CNe may undergo the SSS~phase during outbursts~\citep{HachisuKato2006}. The ejected envelope becomes transparent some time after the outburst maximum, and the hot nuclear burning shell, producing the soft X-ray emission, can be observed~\citep{BodeEvans2008}. This picture is confirmed both by calculations~\citep{Starrfield2004, Soraisam_etal2016} and observations~\citep[see e.g.][]{1999A&A...347L..43K,2001A&A...373..542O, 2003ApJ...594L.127N, Henze_etal2014}. 

The SSS phase was named after the classical super-soft sources, which are a sub-class of cataclysmic variable (CV) stars with high mass-accretion rate~${\sim}10^{-7}\,M_\odot$\,yr$^{-1}$ and (quasi-)steady state thermonuclear burning on the white dwarf~(WD) surface~\citep{van_den_Heuvel_etal1992,Rappaport_etal1994}. CVs are close binaries, consisting of a WD as a primary star and a main-sequence or red giant companion that fills its Roche lobe. As a result, the matter flows from the companion to the WD (with the possible formation of an accretion disc). CNe originate from thermonuclear explosions on WD~surfaces in such objects~\citep{Starrfield_etal2016}. Namely, the thermonuclear explosion takes place in the accumulated hydrogen-rich material on the surface of the WD, which leads to the expansion of the envelope and to subsequent rise in the brightness of the system, as well as to mass ejection from the binary~\citep[see][for a review]{BodeEvans2008}.

The first SSSs were discovered in the Magellanic Clouds by the \textit{Einstein} observatory~\citep{Long_etal1981, Seward_Mitchell1981} and then were treated as a new class of X-ray objects during the ROSAT observations~\citep{Trumper_etal1991}. They are characterised by very soft thermal X-ray spectra with blackbody temperatures of about $15-80$~eV ($150-900$~kK) and high luminosities ${\sim} 10^{36}-10^{38} \rm\, erg/s$, comparable with the Eddington luminosity of a solar mass object~\citep{Kahabka_van_den_Heuvel1997, Greiner2000}. Currently more than $100$~SSS are known in ${\sim}20$ external galaxies, the Magellanic Clouds and our Galaxy~\citep{Greiner2000, Kahabka2006, Maitra_etal2022}.

Interpretation of SSS spectra is challenging. Low-resolution spectra can be often reproduced by a blackbody fit, but it is not a physically correct model. In general, blackbody fitting overestimates the interstellar neutral hydrogen column density~$\nh$ and underestimates temperature, resulting in an overestimated luminosity~\citep[see e.g.][]{Greiner_etal1991, Heise_etal1994}. High-resolution spectra of SSSs are much more complex, with plenty of emission and absorption lines \citep[with the clear role of orientation effects, see][]{Ness_etal2013}. 

Physically motivated models, which potentially can describe the observed SSS spectra are the theoretical spectra of hot WD~atmospheres. \citet{Heise_etal1994} constructed the first model atmospheres of hot~WDs for explaining the SSS X-ray spectra. These models were computed under the assumption of local thermodynamic equilibrium~(LTE). However, deviations from this assumption potentially can be significant at high temperature. Therefore, various authors tried to use the model spectra of hot WD~atmospheres computed without LTE assumption (non-LTE, NLTE model atmospheres). Particularly, the \textsc{tlusty}~\citep{HubenyLanz1995, TLUSTY_ascl} code was used to model hot WD atmospheres and corresponding model spectra were used for fitting the BeppoSAX and ROSAT spectra of various SSSs \citep[e.g.][]{HartmannHeise1996, HartmannHeise1997, Hartmann_etal1999, Parmar_etal1997, Parmar_etal1998}. \citet{Lanz_etal2005} performed a detailed NLTE~model atmosphere analysis of \textit{Chandra} and \textit{XMM-Newton} spectroscopy of the SSS CAL~83 in the LMC.

The multi-purpose model-atmosphere NLTE~code \textsc{phoenix}~\citep{HauschildtBaron1999, PHOENIX_ascl}, modified to calculate model atmospheres for the late X-ray phase of CNe, was used to fit the \textit{Chandra} observations of nova V4743~Sgr~\citep{Petz_etal2005}. The WD~atmosphere was 1D~spherical and was approximated by an expanding, but stationary in time, structure. The composition was assumed to be solar, leading to some uncertainties in the fit. Due to the strong evidence for expansion in the SSS spectra of novae~\citep{Ness2010}, a new version of \textsc{phoenix} was developed to account for this wind-type ejecta expansion~\citep{vanRossumNess2010, vanRossum2012}. Also, a four-dimensional grid of wind-type model spectra was introduced for $\Teff = 450-750 \,\rm kK$, $7$~values of $\log g$, $4$~values of the wind asymptotic velocity~$v_{\infty}$ and $3$~values of mass-loss rate~$\dot{M}$, while the chemical composition was set to solar~\citep{vanRossum2012}.

The T\"ubingen NLTE Model-Atmosphere Package \citep[\textsc{tmap}, ][]{Rauch2003, Werner2003, TMAP_ascl} 
model atmospheres were successfully used for spectral analysis of hot, compact stars \citep[see][for a brief summary]{RauchWerner2010}. Using this code, \citet{Rauch_etal2010} performed a NLTE spectral analysis of the nova V4743~Sgr in SSS~phase. While the model was plane-parallel and static (in contrast with \citealt{Petz_etal2005}), the chemical composition was more realistic. Additionally, the grid of models was calculated for \element{H}-\element{Ni} with solar abundance ratios within $\Teff = 450-1050 \,\rm kK$ and a fixed surface gravity of $\log g = 9$~\citep[see refs. in][]{Rauch_etal2010}.

The challenging factor in describing spectra of SSSs is the unknown composition of the WD atmosphere, which should reflect both the composition of the accreted material and the nucleosynthesis products. Therefore, a mixture of stellar and CNO-cycle processed material is expected. The resulting deficiency of hydrogen and enhancement of helium abundances, as well as the deficiency of carbon and the enhancement of nitrogen and oxygen abundances, have an impact on the strength of absorption lines and the general continuum shape~\citep[e.g.][]{Rauch2003, Rauch_etal2010}.

Nevertheless, LTE model atmospheres were used by \citet{Ibragimov_etal2003} and \citet{Suleimanov_Igragimov2003} for the analysis of ROSAT spectra of $11$~SSSs. They used  a modified version of Kurucz's \textsc{atlas}~code. This models were also used to fit spectra of other SSSs~\citep[]{Burwitz_etal2007, Swartz_etal2002}.

Recently \citet{Suleimanov_etal2024} further refined the LTE~approach, including a more extensive set of spectral lines and photoionisation opacities from the excited levels of heavy element ions. They successfully applied a computed set of model atmosphere spectra to the interpretation of grating \textit{Chandra} spectra of two classical SSS~sources, CAL~$83$ and RX~J$0513.9-6951$. It was shown that the new LTE~model atmospheres allow to represent the grating \textit{Chandra} spectrum of CAL~$83$ with the same accuracy as the non-LTE atmospheres computed by \citet{Lanz_etal2005} and provide similar results. But the LTE~approach has several advantages over~NLTE, with the main advantage being faster computational speed, allowing to compute much more models with various parameters. Additionally, it is easier to take into account a large number of ions and their excited levels as well as  a larger number of spectral lines compared to the NLTE case.

In this article we present a new grid of high-gravity hot LTE~model atmospheres with various chemical compositions, computed using the code developed by \citet{Suleimanov_etal2024}. This grid has been used to fit eROSITA and {\it XMM-Newton} spectral observations of the supersoft phase of the CN~AT~2018bej and we present here the obtained results.

The used X-ray observations are described in Sect.~\ref{sect:observations}. The model atmospheres and the corresponding grid of emergent spectra computed for fitting the observed spectra are presented in Sect.~\ref{sect:model_atmospheres}. The results of spectral fitting are presented in Sect.~\ref{sect:analysis}, while in Sect.~\ref{sect:discussion} we discuss it in the context of modern models of thermonuclear-burned envelopes of WDs. We summarise the obtained results in Sect.~\ref{sect:summary}.

\section{Observations}
\label{sect:observations}

\begin{table*}
    \caption{X-ray observation log of AT~2018bej.} 
    \label{tab:log}
\begin{center}
    \begin{tabular}{lccccc}\hline\hline \noalign{\smallskip}
        Telescope & ObsID & Date (UT) & Date (MJD) & Exp. time$^b$ & Count rate$^c$ \\[0.3em] \hline\noalign{\smallskip}
        eROSITA/eRASS1 && 12--29 Feb. 2020 & 58900.12$^a$ & $3.96$ & $16.4$ \\
        eROSITA/eRASS2 && 16 Aug. -- 2 Sep. 2020 & 59085.75$^a$ & $4.04$ & $12.6$ \\
        eROSITA/eRASS3 && 31 Jan. -- 12 Feb. 2021 & 59251.60$^a$ & $3.17$ & $0.52$ \\
        XMM/RGS & $0854591001$ & 14 Mar. 2020 & $58922.02$ & $84.8/83.6^d$ & $0.44/0.48^d$ \\
        \hline 
    \end{tabular}
\end{center}
Notes: (a)~-- MJD of the midpoint of the observation; (b)~-- The corrected net exposure time in ks; (c)~-- mean count rate in counts per sec; (d)~-- Exposure time and count rate of RGS1 and RGS2 instrument, respectively.
\end{table*}

\subsection{eROSITA}

As part of the eRASS1 all-sky survey, eROSITA onboard the Spectrum-Roentgen-Gamma (SRG) observatory \citep{Predehl_etal2021} started scanning over the position of AT~2018bej on February~12,~2020, with visits every $4$ hours of $50$-seconds duration. A new X-ray source was found at RA(J2000) = $06^{\rm h}26^{\rm m}21^{\rm s}.58$ Dec(J2000) = $-69^{\circ}45'49''.5$ with an estimated uncertainty of $10\arcsec$ radius (including systematics). This position is consistent with the known CV~AT~2018bej/ASAS-SN~18jj~\citep{telegram_xray, telegram_optic}. Later this region was scanned during eRASS2--eRASS4 surveys in 2020 and 2021, see the observation log in Table~\ref{tab:log}.

The data was reduced with the task \texttt{srctool} from the eROSITA Standard Analysis Software System~\citep[\texttt{eSASS,}][]{Brunner_etal2022} version~211214, pipeline configuration~\texttt{c020}~\citep[][see Table E.1]{Merloni_etal2024}. We defined an annulus region with $20\arcsec$~and $100\arcsec$~inner and outer radius, respectively, to extract the source from eRASS1--eRASS2 observations and an inner circle with 31\arcsec~radius in the case of eRASS3, but the source was barely visible during the eRASS3 survey, and it was not visible in eRASS4, so the eRASS4 observation is not shown in Table~\ref{tab:log}. For background extraction we used a 130\arcsec~circle from a nearby source-free region.

\begin{figure*}
    \centering
    \includegraphics[width=1.0\linewidth]{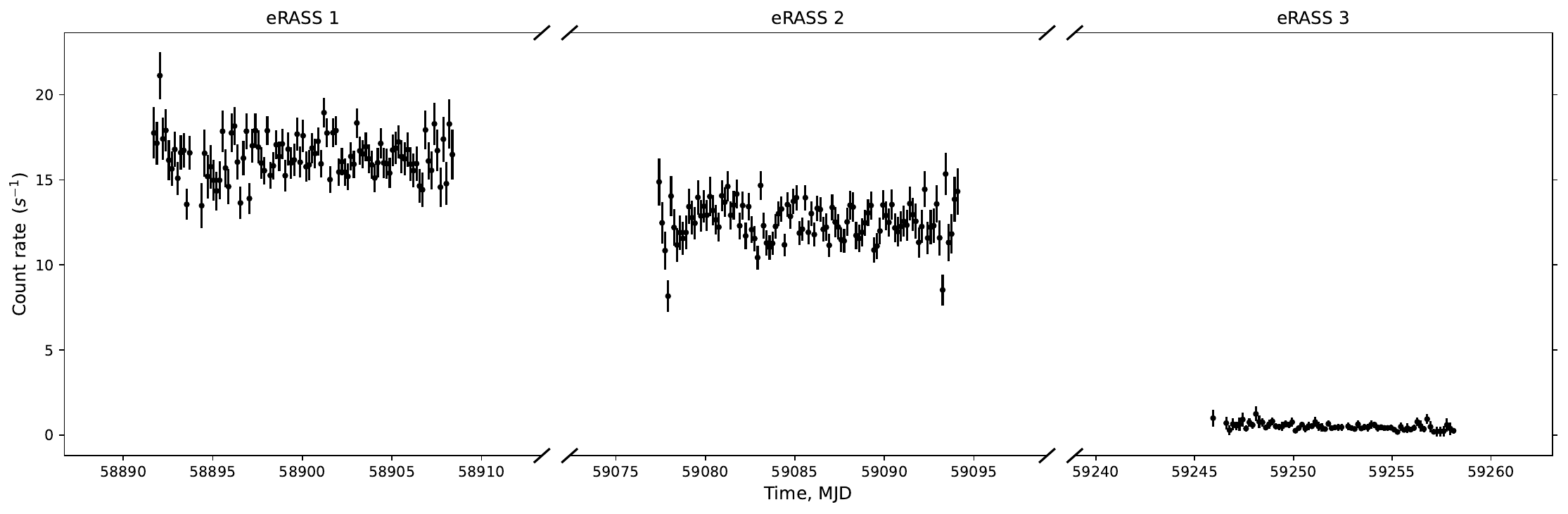}
    \caption{eROSITA light curve of AT~2018bej in the energy band $0.2-2$~keV, combining data from TM1--TM4, and TM6. eROSITA scanned the source position during four epochs eRASS1--4 until the middle of 2021. The source was barely visible during eRASS3~(right panel) and was not detected during eRASS4. The {\it XMM-Newton} observation started ${\sim}10$ days after the end of eRASS1 scanning. The~\citet{Gehrels1986} approximation is used for bins with unknown count rate errors.}
    \label{fig:eROSITA_lc}
\end{figure*}

\begin{figure}
    \center{\includegraphics[width=1.0\linewidth]{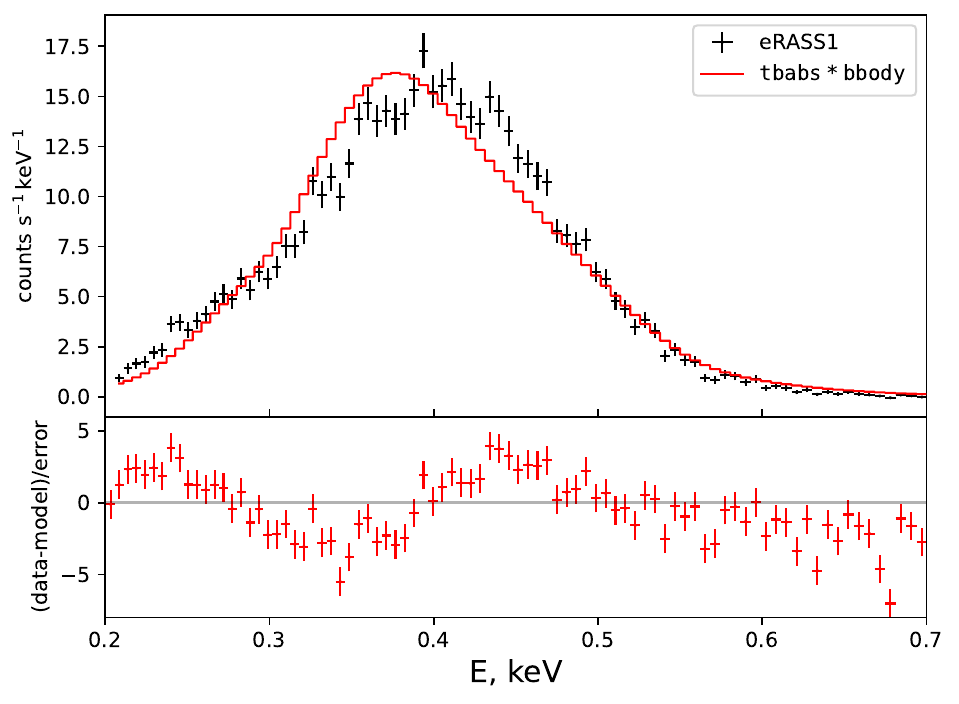}}
    \caption{The eROSITA eRASS1~spectrum, combining data from TM1--TM4, and TM6. Also shown is the absorbed blackbody model. The obtained blackbody parameters are presented in Table~\ref{tab:bb_fit}.}
    \label{fig:spectrum_eROSITA_bb}
\end{figure}

We combined the data from TM1–-TM4 and TM6, the five cameras with an on-chip filter. We did not use data from TM5 and TM7 as they have calibration problems due to a light leak~\citep{Predehl_etal2021}. A cut in the fractional exposure of $0.15$ ($\texttt{FRACEXP} > 0.15$) was applied to light curves to exclude data from the edge of the detectors. The obtained light curves are shown in Fig.~\ref{fig:eROSITA_lc}, while the spectrum (eRASS1 epoch) is shown in Fig.~\ref{fig:spectrum_eROSITA_bb}. The eRASS1 light curve is constant on long time scales, indicating that the supersoft state started prior to the beginning of the eROSITA observations.

\subsection{XMM-Newton}

\begin{figure}
    \centering
    \includegraphics[width=1.0\linewidth]{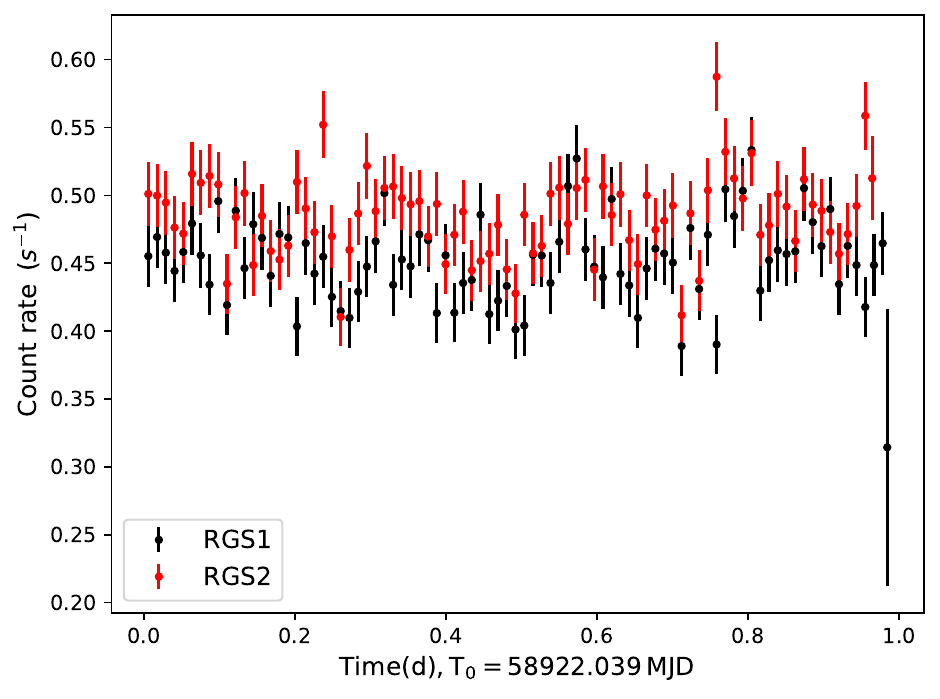}
    \caption{{\it XMM-Newton} light curve of AT~2018bej in the energy band $0.3-3$~keV from both RGS1 and RGS2 instruments.}
    \label{fig:XMM_lc}
\end{figure}

Following the eROSITA discovery, a proposal for high-resolution spectral observation was made at the {\it XMM-Newton} telescope. The source was observed by {\it XMM-Newton} Reflection Grating Spectrometer \citep[RGS,][]{denHerder_etal2001} instrument on~2020~March~14, 10~days after the last eRASS1 observation, for the observation log see Table~\ref{tab:log}. The data was processed by the {\it XMM-Newton} Science Analysis Software (\texttt{SAS~19.0.0}) package; the first-order spectra were used. The obtained light curve is shown in Fig.~\ref{fig:XMM_lc}. It also does not show any obvious periodicity.

\subsection{Preliminary analysis}

\begin{table}
\begin{center}
    \caption{Preliminary fit of AT~2018bej with blackbody model.}
    \label{tab:bb_fit}
    \begin{tabular}{lcc}
    \hline \hline \\ [-0.5em] 
    & eRASS1 & XMM  \\ [0.3em]
    \hline \\ [-0.7em] 
    $\nh, 10^{20}\rm\,cm^{-2}$  & $14.2\pm 0.5$ &  $23.3\pm 0.3$  \\ [0.3em]
    $\Teff,\,\rm kK$            & $423\pm 8$  &  $346\pm 2$  \\ [0.3em]
    $R^a,\,\rm 10^8 cm$        & $103\pm_{13}^{15}$ &  ${\sim}\,600$  \\ [0.3em]
    $L, \,10^{37}\,\rm erg\,s^{-1}$   & $240\pm_{60}^{70}$  &  $3700\pm 100$ \\ [0.3em]
    ${\tt cstat}\,{\rm (dof)}$ & $412.90\, (128)$ &  $5481.75\, (3325)$ \\ [0.1em]
    \hline
    \end{tabular}
\end{center}
Notes: (a)~-- the distance to the~LMC is assumed to be~$50\,\rm kpc$~\citep{Pietrzynski_etal2019}.
\end{table}

The Bayesian approach was used for analysing the X-ray spectra. We applied the analysis software Bayesian X-ray Analysis~\citep[BXA,][]{Buchner_etal2014, BXA_ascl}, which connects the nested sampling package UltraNest\footnote{\url{https://johannesbuchner.github.io/UltraNest/}}~\citep[][]{Buchner2021, UltraNest_ascl} with~\textsc{xspec}\,\footnote{\url{https://heasarc.gsfc.nasa.gov/docs/xanadu/xspec/}}\citep{Arnaud1996, XSPEC_ascl}. Posterior probability distributions and the Bayesian evidence were derived with the nested sampling Monte Carlo algorithm MLFriends~\citep{Buchner_etal2014, Buchner2019}. To construct the corner plots, the \texttt{corner}\footnote{\url{https://corner.readthedocs.io/en/latest}} package was used~\citep{Foreman-Mackey2016, corner_ascl}.

It is known that for low-count spectra the usage of C-statistics~\citep{Cash1979} is preferable instead of the $\chi^2$-statistics. We used its~\textsc{xspec} implementation~\texttt{cstat} as a likelihood in order to determine the best-fit parameters. The parameter uncertainties are derived from the~$0.16$~and~$0.84$~quantiles of the posterior distribution (which corresponds to $68\%$~CL). As there is no convenient criterion of estimating the goodness-of-fit (like $\chi^2_{\nu}$) for~\texttt{cstat}, we will give below the statistics value and the number of degrees of freedom.

At first, the eROSITA and RGS~spectra were fitted by the simple blackbody model, while for absorption we used the T\"ubingen-Boulder ISM absorption model {\bf\tt tbabs}. The eROSITA spectrum with the blackbody fit is shown in Fig.~\ref{fig:spectrum_eROSITA_bb}.
However, these results are considered non-physical, because the emitting radii of $(1-6)\cdot10^5\, \rm km$ (assuming a LMC distance of $50$~kpc, see Table~\ref{tab:bb_fit}) are inconsistent with a typical WD radius, and the resulting luminosity significantly exceeds the Eddington limit for typical WD~masses. This is not surprising, as it is well known that blackbody fits tend to overestimate~$\nh$ and underestimate the temperature \citep[see e.g.][]{Greiner_etal1991, Heise_etal1994}. Therefore hot WD~model atmosphere spectra are necessary to reproduce SSS~spectra.

Alternatively, \citet{Suleimanov_etal2024} computed a new set of LTE~model-atmosphere spectra and showed that these models allow to represent the spectra of two classical SSS~sources, CAL~$83$ and RX~J$0513.9-6951$. In the cited article, also three model grids were calculated for three chemical compositions: the H/He~mix was solar, and the abundance of heavy elements was set to solar, $0.1*$solar, and $0.5*$solar. The latter abundances correspond to the~SMC~and~LMC~composition. A preliminary analysis of all spectra with these models gives $\Teff \approx 570-600\,\rm kK$ and $\log g\approx 8.25$ 
with the LMC~composition preferred.

\section{Model atmospheres}
\label{sect:model_atmospheres}

For describing the observed X-ray spectrum of the object, we computed our own model-grid spectra of hot WDs. We used our code, which is based on Kurucz's code \textsc{atlas}~\citep{Kurucz1970}, and modified for high temperatures and high densities \citep{Ibragimov_etal2003, Suleimanov_etal2006, Suleimanov_etal2015}. The $15$~most abundant chemical elements from~\element{H} to~\element{Ni} were taken into account and about $20,000$~lines from CHIANTI, Version~3.0, atomic database~\citep{Dere_etal1997}. The main model parameters are the effective temperature $\Teff$ and the gravity $\dlg = \log g - \log g_{\rm Edd}$. The latter indicates the distance of model from the Eddington limit:
\begin{equation}
    \log g_{\rm Edd} = \log(\sigma_{\rm e}\sigma_{\rm SB}T^4 /c) = 4.818 + 4\log(\Teff/10^5 \rm K),
\end{equation}
see details in \citet{Suleimanov_etal2024}. 

The chemical composition of the WD~atmosphere during the SSS~phase after the nova explosion is significantly different from that of the accreted matter \citep[see e.g.][]{Rauch_etal2010} due to thermonuclear fusion products being brought to the surface. Therefore, it is necessary to calculate model atmospheres with chemical compositions close to the expected one.

It was assumed that the initial chemical composition was LMC-like~(solar H/He mix and half-solar heavy elements abundance), then some part of the hydrogen was transformed into helium via the CNO-cycle, resulting in an increase in the number fraction of~CNO. We used the composition of Rauch's NLTE~models as a starting point. The ratio of the number fractions of hydrogen and helium in these models are equal to each other, the carbon number fraction is less than solar, and the number fractions of nitrogen and oxygen are increased in comparison with the solar values. Namely, in Rauch's model~006\footnote{\url{http://astro.uni-tuebingen.de/~rauch/TMAF/flux_HHeCNONeMgSiS_gen.html}}, which we considered as a fiducial model, ${\rm A_{\rm H}/A_{\rm He}}$ = $0.48/0.48$, and the number fraction of carbon is~${\rm A_{\rm C}}=0.39*$solar. We considered various ratios ${\rm A_{\rm H}/A_{\rm He}}$ and finally adopted $0.46/0.54$. Technically, from a statistical point of view, a ratio ${\rm A_{\rm H}/A_{\rm He}}=0.0/1.0$ is slightly more likely, but we considered it less physical, as it is assumed that active hydrogen thermonuclear burning still occurs at the surface of the WD. The number fraction of carbon was considered as an additional free parameter of the model.

Thus, in addition to $\Teff$ and $\dlg$, a new parameter has been introduced~-- the number fraction of carbon~`\chem' in solar units. The number fraction of nitrogen is varied from~$1.89$ to~$0.5$ in solar units depending on~`\chem' in assumption that $(0.5-\chem)$ of carbon is transformed into nitrogen. Thus, the resulting abundance of carbon~`\chem' changes from~$0$ (all C$\rightarrow$N) to~$0.5$ (LMC~composition). The abundance of other elements is assumed to be $0.5*{\rm solar}$ (we use solar abundances from~\citet{Asplund_etal2009}).

Based on the preliminary fit analysis, altogether $288$~models have been calculated for $T_{\rm eff}$ from $525\, \rm kK$ to $700\, \rm kK$ in steps of $25\, \rm kK$, $\dlg=0.1,\, 0.2,\, 0.4,\, 0.6,\, 1.0,\, 1.4,$ and $\chem = 0.0,\, 0.1,\, 0.2,\, 0.3,\, 0.4,\, 0.5$. We will further refer to this model as the \chem-model. All spectra were implemented as a tabular \textsc{xspec}-format model.

\begin{figure}
    \center{\includegraphics[width=1.0\linewidth]{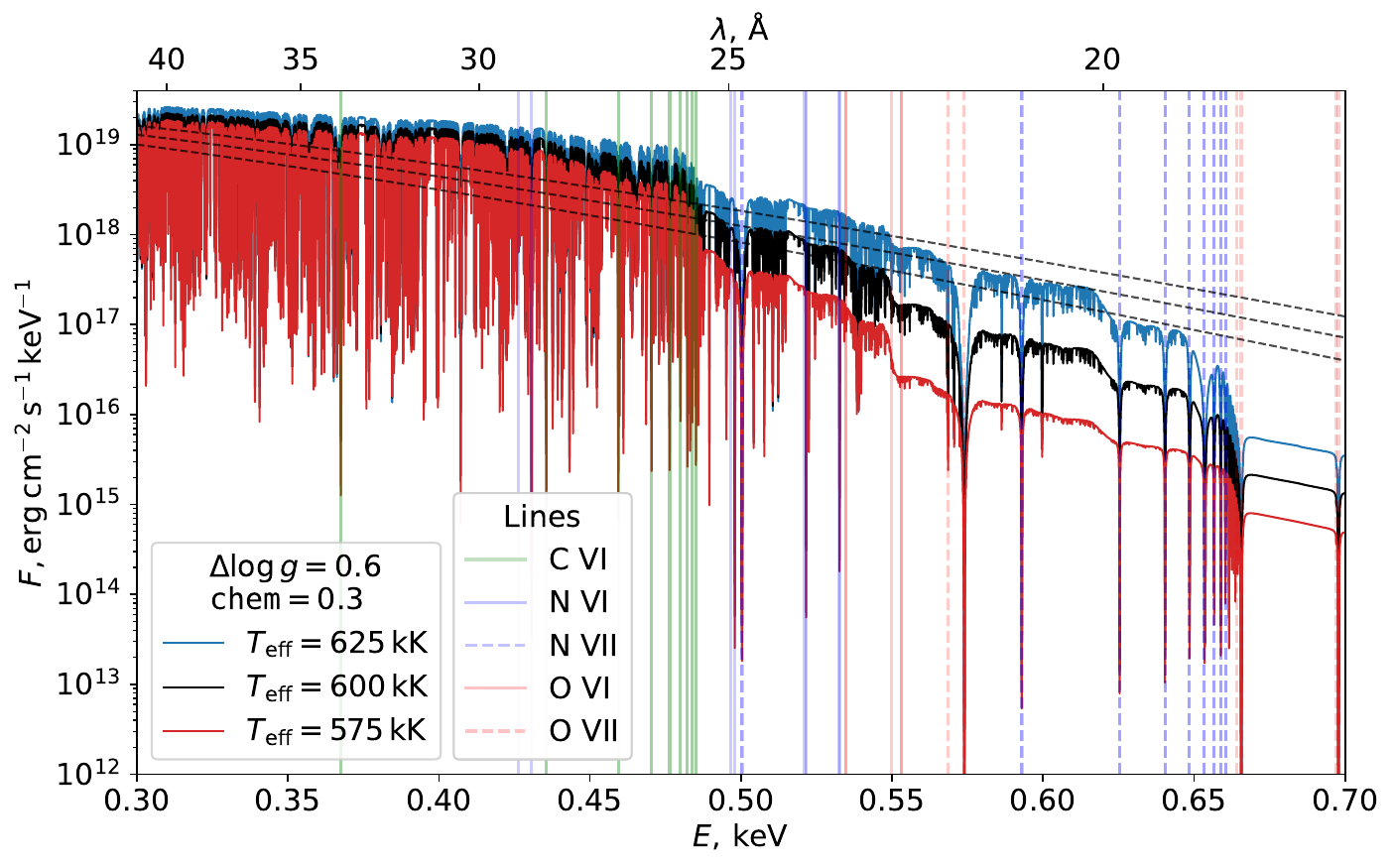}}
    \caption{The emergent spectra of model atmospheres with different $\Teff$. Spectral series for some {\rm CNO}~ions are also indicated. The grey dashed lines represent the corresponding blackbody spectra.}
    \label{fig:diff_teff}
\end{figure}

\begin{figure}
    \center{\includegraphics[width=1.0\linewidth]{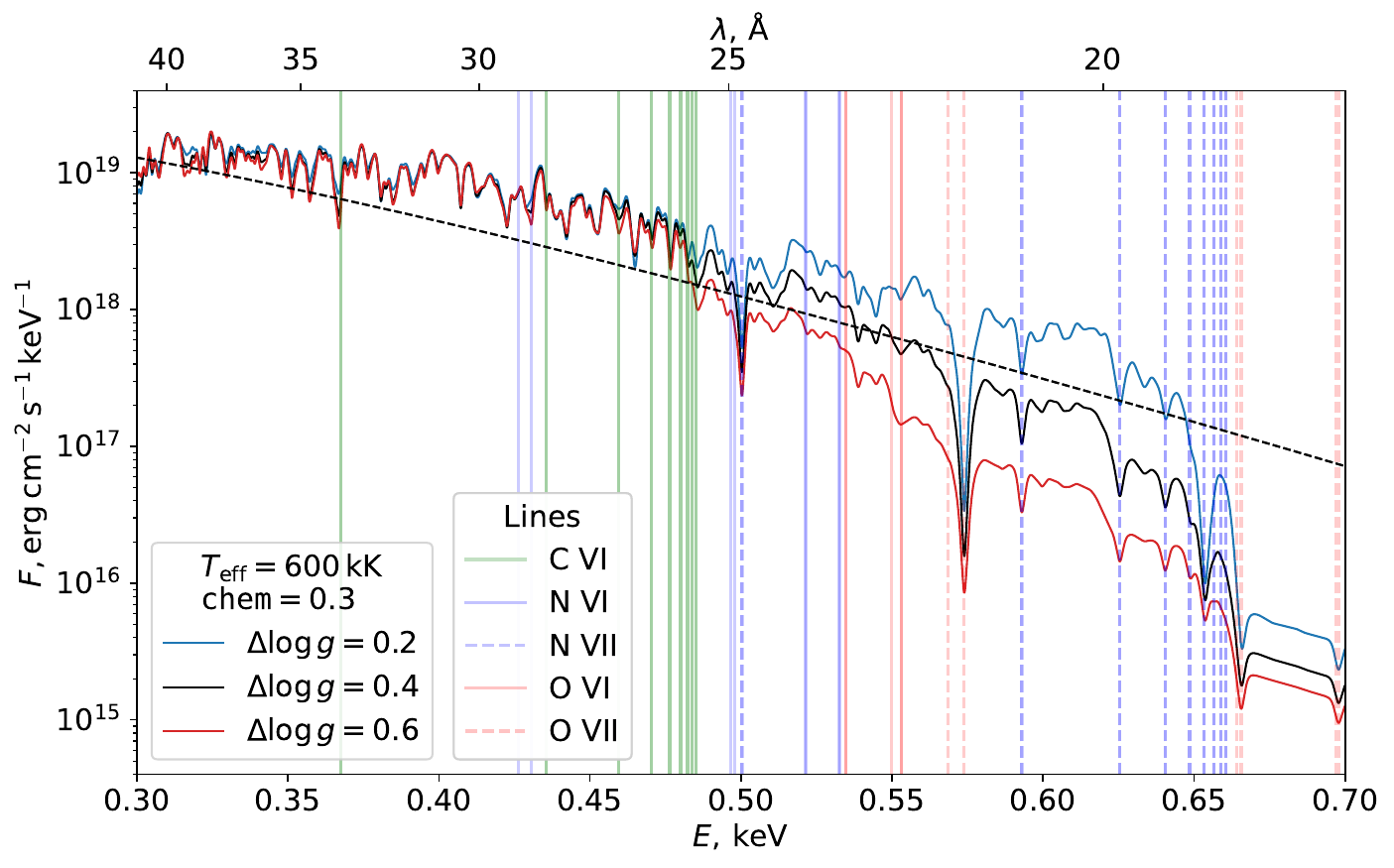}}
    \caption{The emergent model spectra with various surface gravity (three values of gravity~$\dlg$). The spectra are smoothed using a Gaussian kernel with resolution $R\approx650$. Notations are the same as in Fig.~\ref{fig:diff_teff}.}
    \label{fig:diff_dlg}
\end{figure}    

\begin{figure}
    \center{\includegraphics[width=1.0\linewidth]{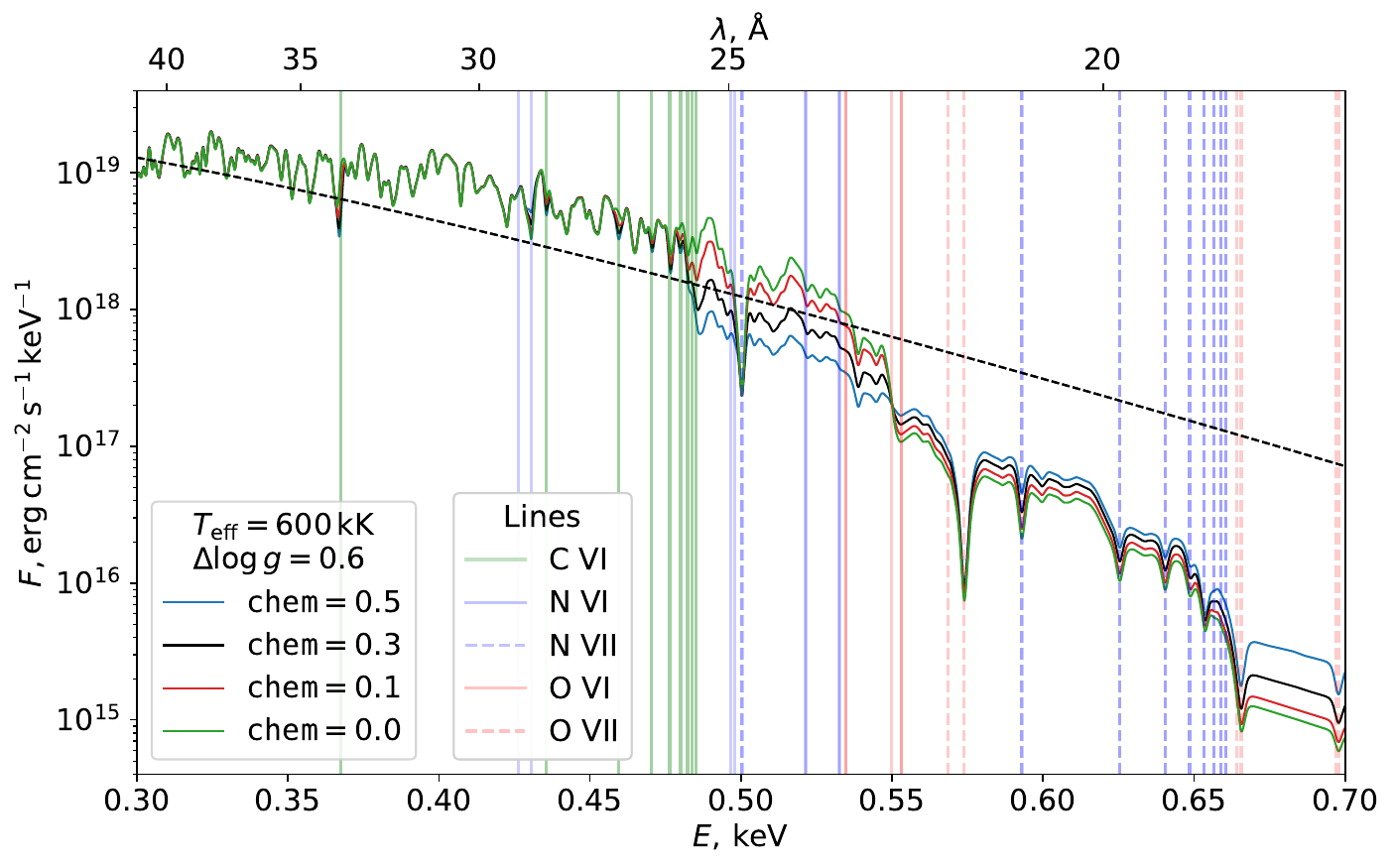}}
    \caption{The emergent model spectra with different values of the \chem~parameter. The spectra are also smoothed~(see Fig.~\ref{fig:diff_dlg}). Notations are the same as in Fig.~\ref{fig:diff_teff}.}
    \label{fig:diff_chemC}
\end{figure}

Examples of emergent spectra of the models are shown in Figs.~\ref{fig:diff_teff} and~\ref{fig:diff_dlg} for different values of~$\Teff$ and~$\dlg$, while $\chem=0.3$ is fixed. The absorption edges corresponding to photoionisation from the ground level of~\ion{C}{VI} ($0.49\,\rm keV$), \ion{N}{VI} ($0.55\,\rm keV$), and~\ion{N}{VII} ($0.66\,\rm keV$) ions are clearly seen. These edges lead to significant deviation of the model atmosphere spectra from the corresponding blackbody flux (indicated by dashed grey lines). The spectra are dominated by a forest of absorption lines (spectral series of some {\rm CNO}~ions are shown), but the main influence on the common shape of the spectra comes from the effective temperature. The spectra in Fig.~\ref{fig:diff_dlg} are convolved for clarity with a Gaussian kernel with {\it XMM-Newton} RGS~resolution $R\approx650$ in the observed soft X-ray energy range. The gravity mainly affects the $0.48-0.65 \rm\, keV$ energy range, which is covered by~RGS.

\begin{figure}
    \center{\includegraphics[width=1.0\linewidth]{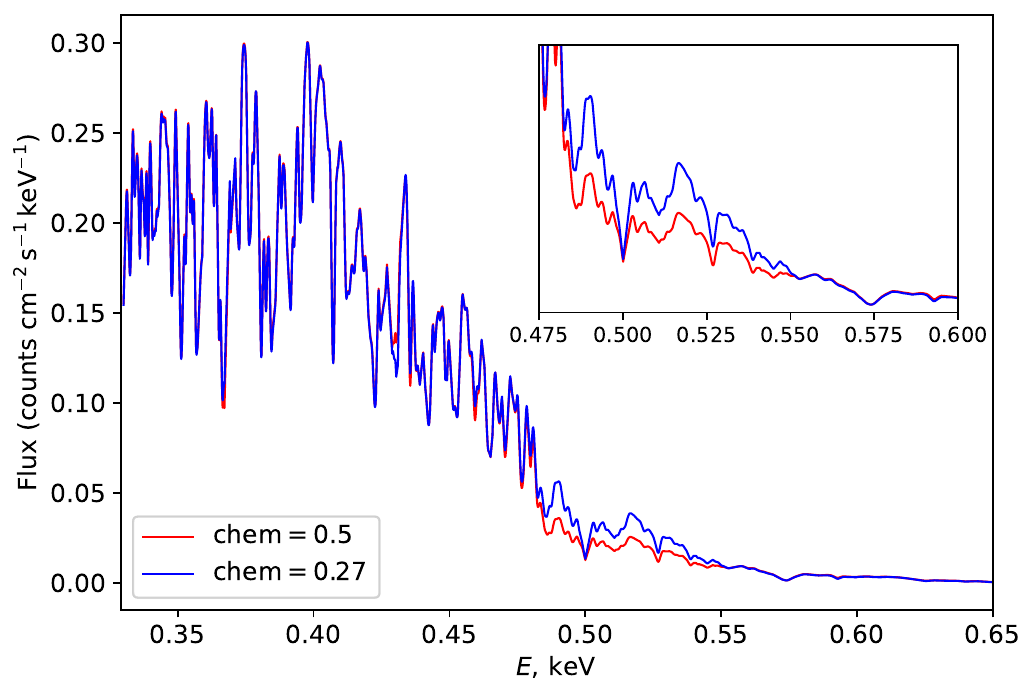}}
    \caption{Comparison of the convolved spectra for different chemical compositions with the same $\Teff=604\,{\rm kK},\, \dlg=0.51,\, {\rm N}_{\rm H} = 5.88\times10^{20}\,\rm cm^{-2}$. Shown are cases for~$\chem=0.0$ (i.e. $\rm A_H/A_{He}=0.46/0.54, A_{\rm rest}=0.5*{\rm solar}$, indicated in blue) and $\chem=0.27$~(the best-fit value of carbon abundance, in red). The~$0.47-0.6\, \rm keV$ region is also shown in detail.}
    \label{fig:diff_LMC_chemC}
\end{figure}

Figure~\ref{fig:diff_chemC} shows some examples of models for different values of the \chem-parameter. The spectra are also convolved with a Gaussian kernel with {\it XMM-Newton} RGS~resolution $R\approx650$ in the observed soft X-ray energy range. The chemical composition mainly affects the $0.48-0.65 \rm\, keV$ energy range, which is covered by~RGS. Although the changes resulting from the chemical composition are smaller, the spectral resolution of {\it XMM-Newton} 
allows for accurate determination, see the convolved spectrum in Fig.~\ref{fig:diff_chemC}. In addition, the overall spectrum level in this region is clearly distinguishable in the~eROSITA spectrum. Also, Fig.~\ref{fig:diff_LMC_chemC} shows the comparison between spectra convolved with the XMM~response for $\chem=0.27$~and~$0.5$~(LMC composition), while temperature and gravity are fixed at $\Teff=605\, {\rm kK}, \, \dlg=0.52$. The notable deviation in the~$0.48-0.55\,\rm keV$ region is clearly visible.

\begin{figure}
    \center{\includegraphics[width=1.0\linewidth]{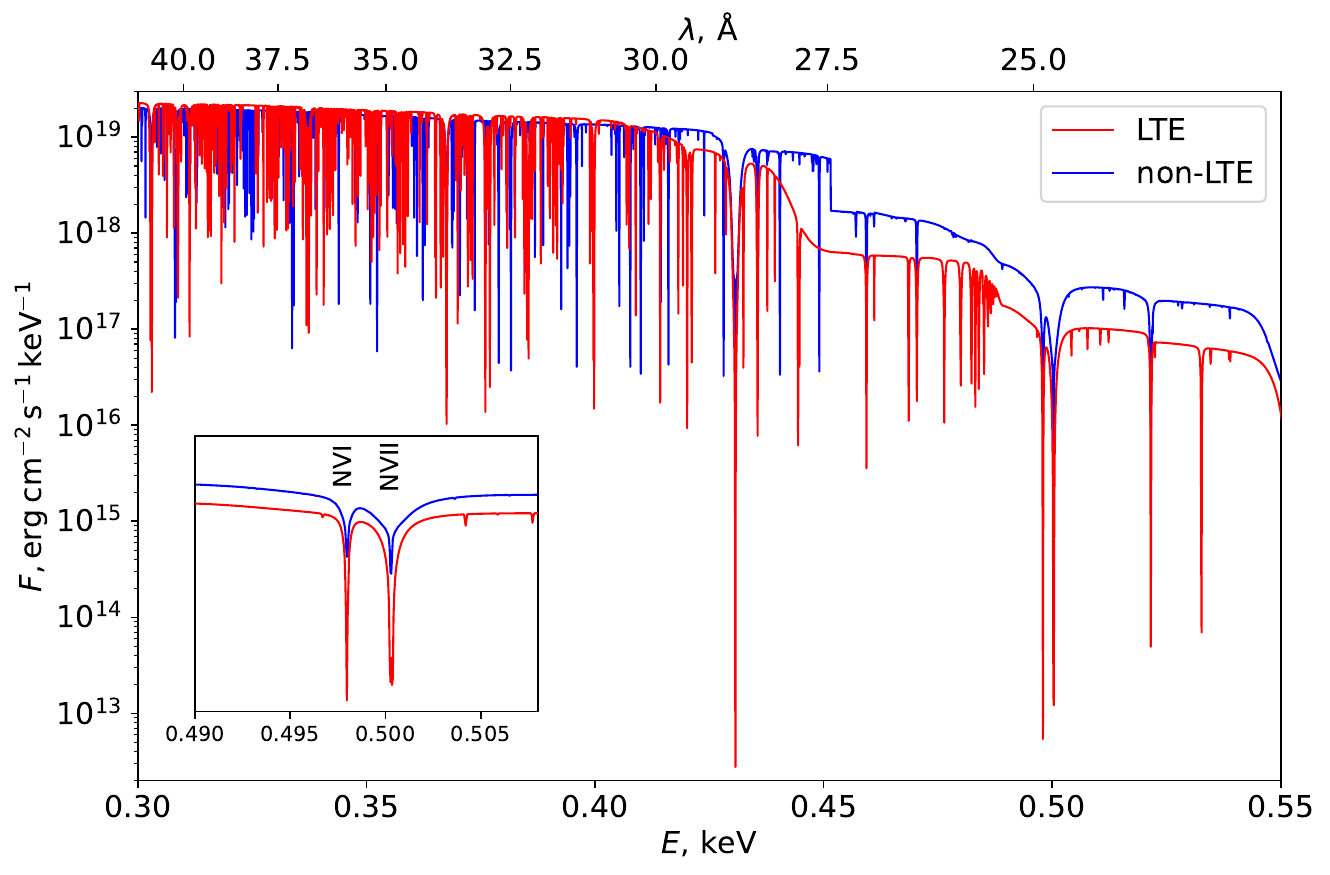}}
    \caption{Comparison of the model atmospheres spectra computed by our code~(in red) and the~\textsc{tmap} non-LTE code~\citep[][in blue]{Rauch_etal2010}.
    Both models have $\Teff = 600\, {\rm kK}, \log g = 9$. The chemical composition corresponds to model~006 from Rauch's grid. The lower left panel shows in detail the comparison of  absorption line profiles.}
    \label{fig:diff_rauch}
\end{figure}

\begin{figure*}
\centering
    \begin{minipage}{0.49\linewidth}
	\center{\includegraphics[width=1.0\linewidth]{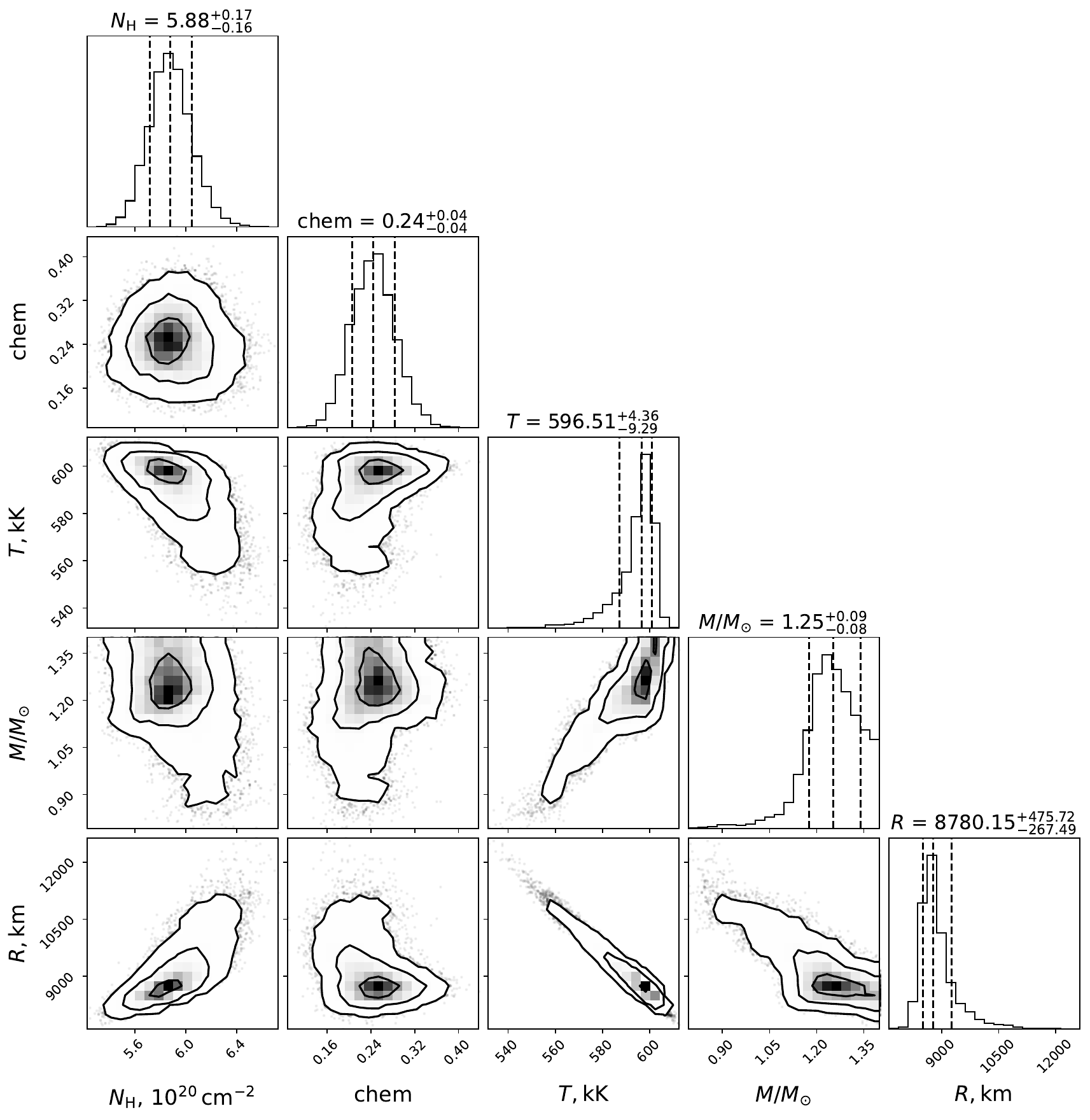}}
	\end{minipage}
	\begin{minipage}{0.49\linewidth}
	\center{\includegraphics[width=1.0\linewidth]{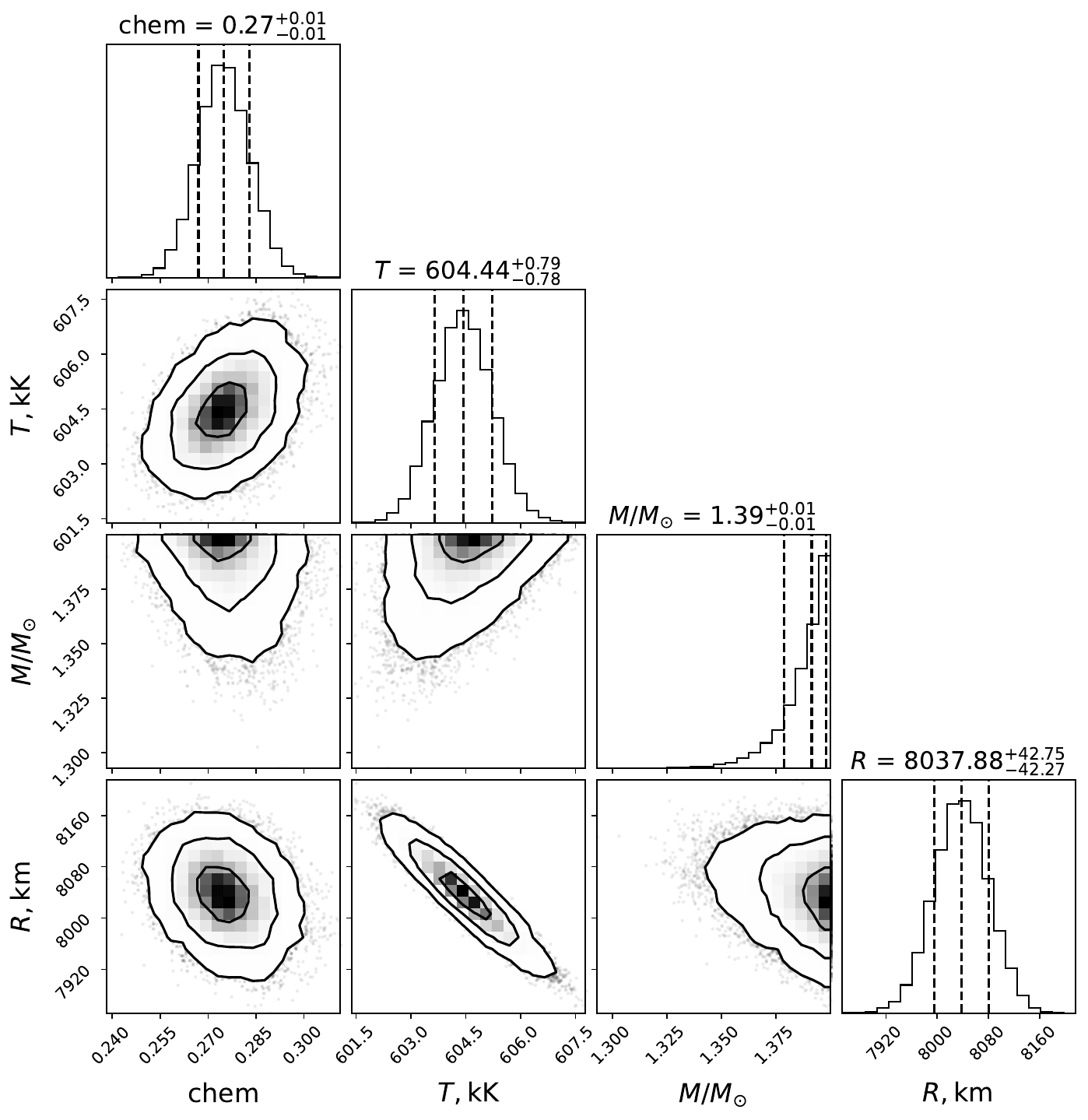}}
	\end{minipage}
    \caption{Corner plot of the parameters' posterior distribution for the eRASS1~spectrum (left panel) and for the {\it XMM-Newton} RGS~spectrum (right panel)  of~AT~2018bej. The WD~mass~$M$ and radius~$R$ are considered as free parameters. The two-dimensional contours correspond to~$39.3\%$, $86.5\%$, and~$98.9\%$ confidence levels. The histograms show the normalised one-dimensional distributions for a given parameter derived from the posterior samples. The hydrogen column density is fixed at~$\nh=5.88\times 10^{20}$~cm$^{-2}$
    for the RGS~spectrum.The best-fit parameter values are presented in Table~\ref{tab:fit}.}
    \label{fig:corner_plot_eRASS1}
\end{figure*}

We also compared our LTE models with NLTE models computed by~\citet{Rauch_etal2010}. Figure~\ref{fig:diff_rauch} shows the comparison between the emergent spectrum of Rauch's NLTE model~006 and our LTE~spectrum with the same chemical composition. Both spectra were binned to a~$0.002\AA$ interval within the~$10 - 60\AA$ wavelength range. The general shapes of the spectra are similar, however, there are some differences in the absorption-edge regions. Due to over-ionisation of the ions~(relative to the LTE~case), the absorption jumps in the NLTE~spectrum appear shallower. Additionally, a sharp jump occurs in the NLTE~spectrum at~$0.45\,\rm keV$, associated with ionisation from an excited level of~\ion{Si}{XI}. This feature is much smoother in our spectrum. Our list of lines is also different, resulting in variations in the line distribution in the spectra. The line broadening of H-like ions seems to differ, as seen in the lower left panel in Fig.~\ref{fig:diff_rauch}, where~\ion{N}{VI} and~\ion{N}{VII} lines are depicted. The broadening of the resonance lines is primarily determined by radiative damping in addition to Doppler broadening. In contrast, the wide wings of the resonance line~\ion{N}{VII} in the NLTE~spectrum are also affected by the Stark effect. Additionally, the level populations differ between LTE~and NLTE~models, causing the lines in the LTE~spectrum to appear deeper. Namely, the upper layers of LTE~models are obviously cooler than those of NLTE~models, and the corresponding ions are more populated. This indicates that a decreased abundance of some elements is needed compared to NLTE~models to describe the same observed spectral line absorption. 

We investigated the uncertainty that may arise from using LTE models instead of NLTE models. It is possible to demonstrate that a NLTE model spectrum with~$T_{\rm eff}=550\,\rm kK$ reproduces almost the same general spectrum shape with the same absorption jump strengths as the LTE~model spectrum with~$T_{\rm eff}=600\,\rm kK$, while the gravity~$\log g$ is fixed, see Fig.~\ref{fig:diff_rauch_550_600}. Reducing~$\log g$ while fixing the temperature leads to the same effect. Thus, using LTE~atmospheres may lead to an overestimation of temperature by~${\sim}\,10\%$ and/or to an underestimation of~$\log g$ by ${\sim}\,0.5$~dex, see Fig.~\ref{fig:diff_rauch_85_9}. 
Note that both of these effects occur simultaneously, and we can only conclude that LTE~model atmospheres slightly overestimate the effective temperature and slightly underestimate the surface gravity of a hot~WD as a result of fitting its soft X-ray spectrum.

This occurs because the elements under NLTE conditions are over-ionised compared to the LTE~case. Consequently, higher temperature and/or less plasma density are needed in LTE~atmospheres to compensate this effect.

\section{Analysis of the spectra}
\label{sect:analysis}

All the observed spectra, eRASS1, eRASS2, and \textit{XMM-Newton}~(see Table~\ref{tab:log}), were fitted with the \chem-model described above. 
Using temperature~$\Teff$ and gravity~$\dlg$ as free parameters leads to a degeneracy in the gravity distribution and to a nonphysical estimation of the WD mass. Therefore, we
fitted the spectra using WD~mass~$M$ and radius~$R$ as free parameters, and set a strict upper limit of~$M=1.4\, M_\sun$. For both eROSITA and XMM~spectra a uniform prior distribution was set for $\nh$ (in the range $(1-10)\times 10^{20} \,{\rm cm^{-2}}$), `\chem'~parameter, effective temperature $\Teff$, WD mass~$M$ (in range $0.3-1.4\, M_\sun$) and radius~$R$ (in range $(2-20)\times 10^3\, \rm km$). Also the additional lower limit was set for the radius to obtain a physically adequate result~-- $R$~should be greater than the radius of cold WD $R_{\rm cold}$~\citep{Nauenberg1972} with the same mass.

\begin{figure}
\centering
	\center{\includegraphics[width=1.0\linewidth]{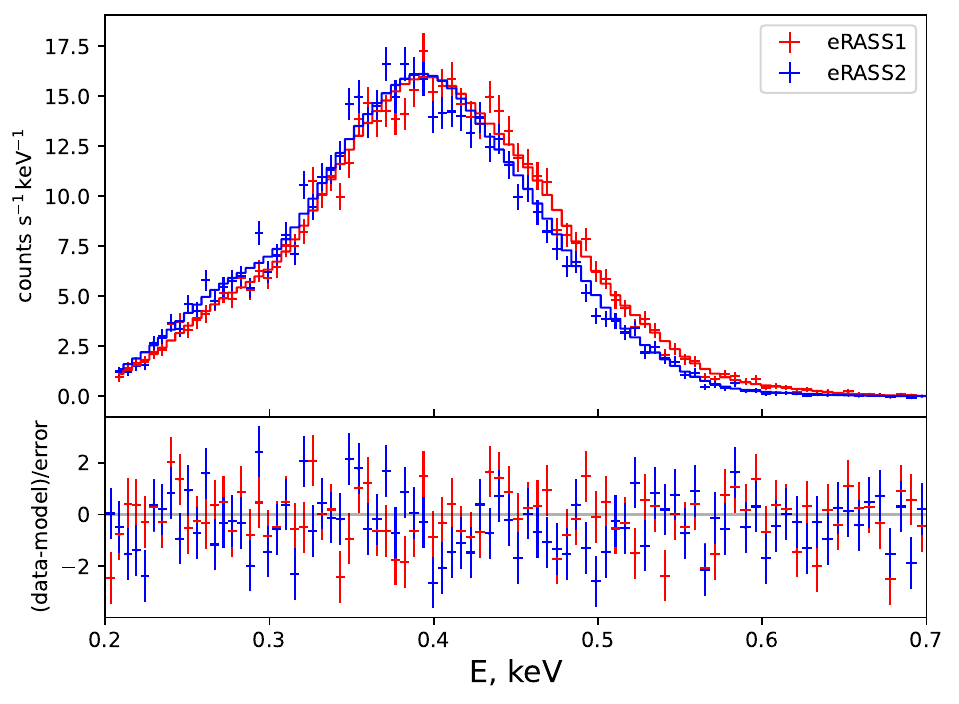}}
    \caption{Model atmosphere best-fit of the eROSITA eRASS1~(in red) and eRASS2~(in blue) spectra. The obtained parameters are presented in Table~{\ref{tab:fit}}.}
    \label{fig:spectrum_eROSITA_our}
\end{figure}

\begin{figure*}
\centering
    \begin{minipage}{0.49\linewidth}
	\center{\includegraphics[width=1.0\linewidth]{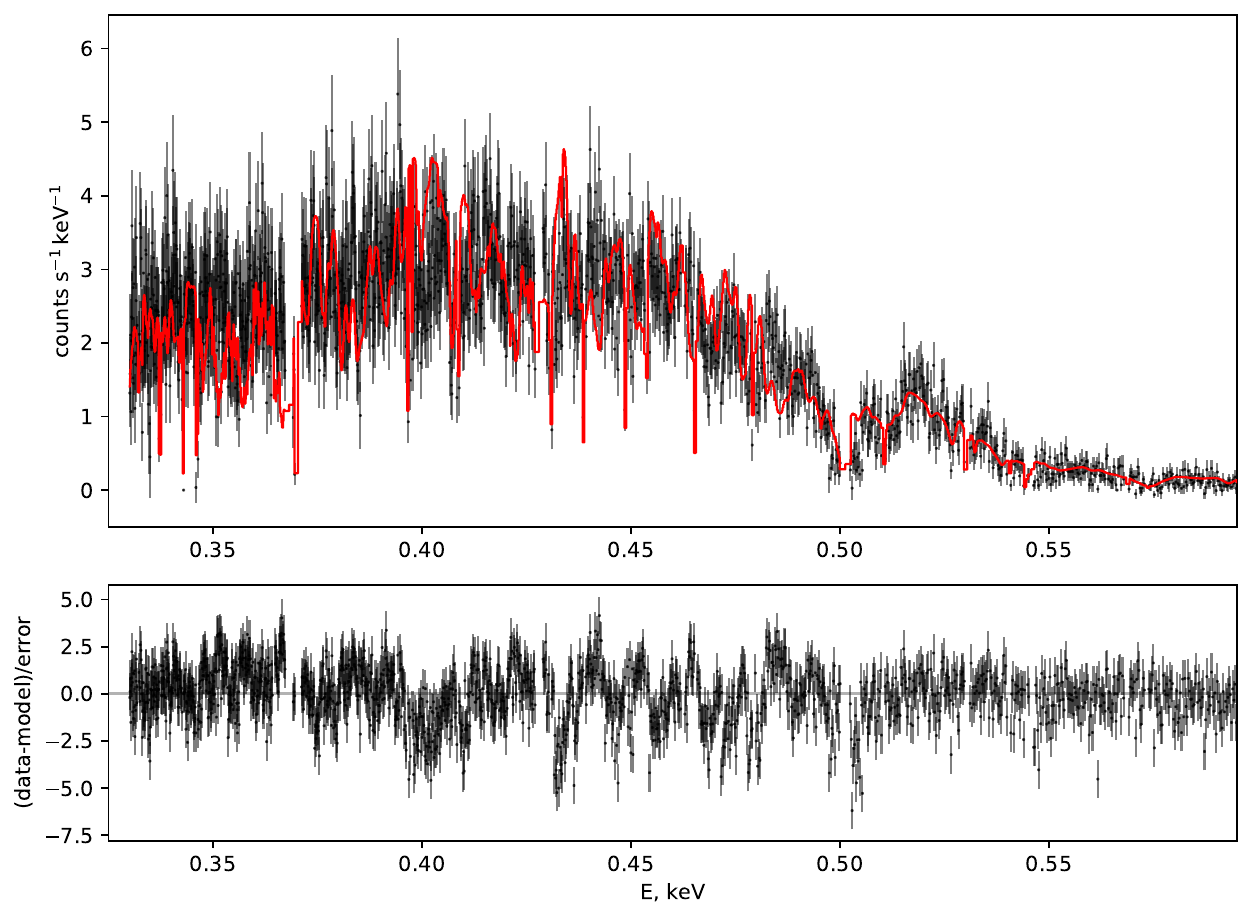}}
	\end{minipage}
	\begin{minipage}{0.49\linewidth}
	\center{\includegraphics[width=1.0\linewidth]{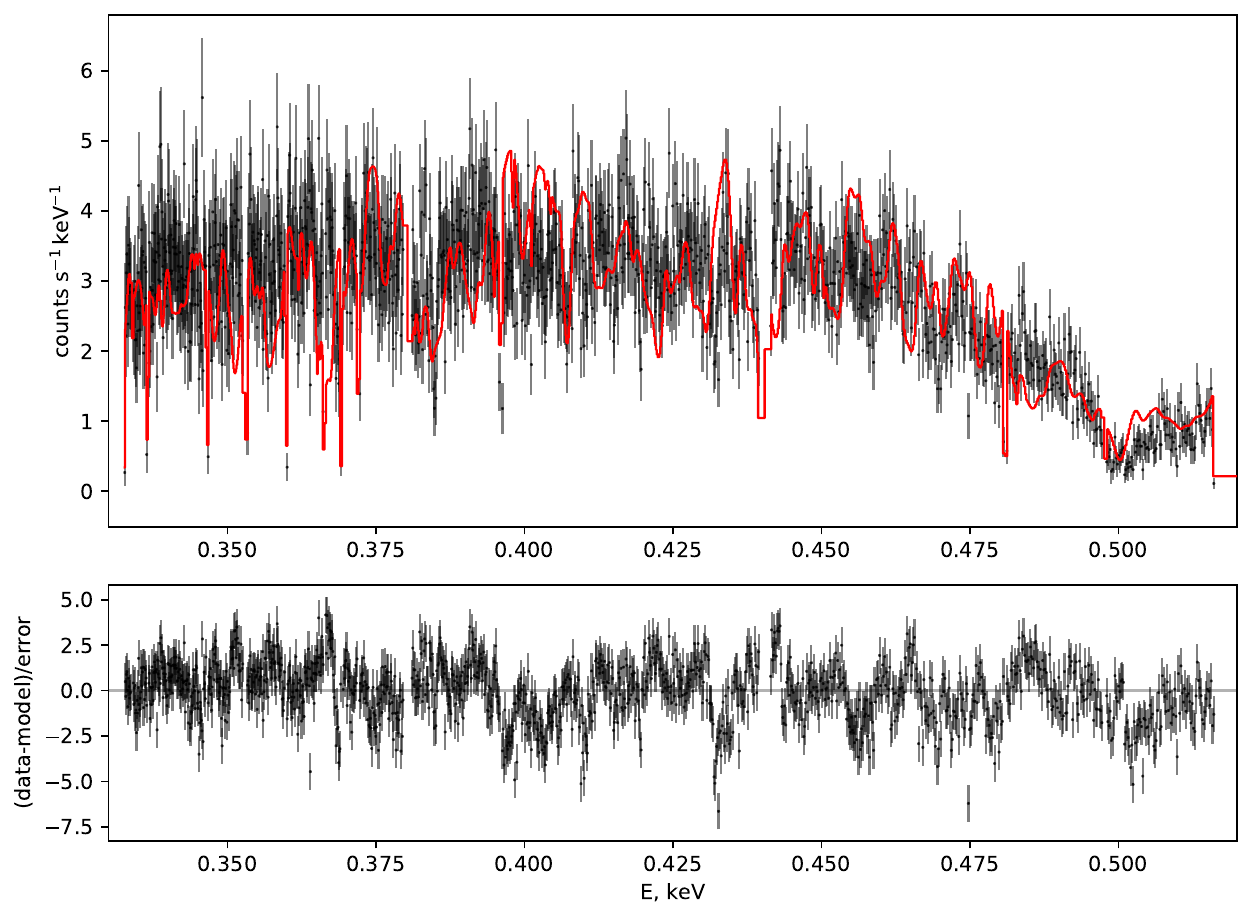}}
	\end{minipage}
    \caption{The {\it XMM-Newton} RGS spectrum together with best-fit LTE model. Shown are spectra of RGS~1 (left panel) and RGS~2 (right panel) instruments. The best-fit parameters can be found in Table~\ref{tab:fit}. The hydrogen column density is fixed at~$\nh=5.88\times 10^{20}\,\rm cm^{-2}$.
    }
    \label{fig:spectrum_RGS}
\end{figure*}

The example of the resulting corner plot for the eRASS1 spectrum is shown in the left panel of Fig.~\ref{fig:corner_plot_eRASS1}, while the obtained parameters for all the fits are presented in Table~\ref{tab:fit}. Fig.~\ref{fig:spectrum_eROSITA_our} shows the comparison between the observed and model eROSITA spectrum. The parameters' posterior distributions for all spectra are  narrow enough and have only one mode, resulting in smaller errors. However, the WD~mass tends to its upper limit, especially for the XMM~case.

\begin{table*}
\begin{center}
    \caption{Spectral parameters of the LTE~model fit for AT~2018bej.}
    \label{tab:fit}
    \begin{tabular}{lcccc|cc}
    \hline \hline \\ [-0.5em] 
     & eRASS1 & eRASS2 & XMM & XMM$^a$ & eRASS1$^d$ & eRASS2$^d$ \\ [0.3em]
    \hline \\ [-0.7em] 
    $\nh, 10^{20}\rm\,cm^{-2}$ & $5.88 \pm_{0.16}^{0.17}$  & $5.69 \pm 0.15$ &  $3.06 \pm_{0.13}^{0.12}$ &  $5.88^a$ &  \multicolumn{2}{c}{$5.83 \pm_{0.16}^{0.14}$} \\ [0.3em]
    $\Teff,\,\rm kK$           & $597 \pm_{9}^{4}$  & $541 \pm_{14}^{10}$ &  $631 \pm 1$  & $604 \pm 1$  & $578 \pm_{11}^{14}$  & $554 \pm_{11}^{8}$ \\ [0.3em]
    $M/M_\odot$                      & $1.25 \pm_{0.08}^{0.09}$  & $0.81 \pm_{0.08}^{0.13}$ & $1.39\pm 0.01$  & $1.39 \pm 0.01$  & \multicolumn{2}{c}{$1.07 \pm_{0.11}^{0.08}$} \\ [0.3em]
    $R^b,\,\rm km$                     & $8780 \pm_{267}^{476}$  & $11001 \pm_{629}^{799}$ &  $5934 \pm_{77}^{74}$  & $8038 \pm_{42}^{43}$  & $9541 \pm_{750}^{601}$  & $10393 \pm_{514}^{678}$  \\ [0.3em]
    $L, \,10^{37}\,\rm erg\,s^{-1}$  & $7.0 \pm_{0.6}^{0.8}$  & $7.4 \pm_{1.1}^{1.2}$ &  $4.0 \pm 0.1$  & $6.1 \pm 0.1$  & $7.3 \pm_{1.3}^{1.2}$  & $7.3 \pm_{0.9}^{1.0}$ \\ [0.3em]
    $\log g$                         & $8.33 \pm_{0.04}^{0.06}$  & $7.95 \pm_{0.07}^{0.09}$ &  $8.72 \pm 0.01$  & $8.46 \pm 0.01$  & $8.19 \pm_{0.08}^{0.06}$  & $8.12 \pm_{0.06}^{0.07}$ \\ [0.3em]
    $\dlg$                  & $0.41 \pm_{0.05}^{0.06}$  & $0.20 \pm_{0.08}^{0.10}$ &  $0.70 \pm 0.01$  & $0.51 \pm 0.01$  & $0.33 \pm_{0.09}^{0.08}$  & $0.33 \pm 0.07$ \\ [0.3em]
    $\rm A_{\rm C},\, sol$           & $0.24\pm 0.04$ & $0.20 \pm_{0.04}^{0.05}$ & $0.22\pm 0.01$ & $0.27\pm 0.01$ & $0.21 \pm 0.04$  & $0.16 \pm 0.04$ \\ [0.3em]
    $\rm A_{\rm N},\, sol$           & $1.14 \pm 0.11$  & $1.33 \pm_{0.14}^{0.11}$ &  $1.28 \pm 0.03$  & $1.14 \pm 0.03$  & $1.30 \pm 0.11$  & $1.44 \pm 0.11$ \\ [0.3em]
    ${\tt cstat}\, {\rm (dof)^c}$    & $125.00\, (125)$  & $144.74\, (125)$ &  $6548.12\, (3323)$  & $7019.75\, (3324)$  & \multicolumn{2}{c}{$272.91\, (252)$}  \\ [0.1em]
    \hline
    \end{tabular}
\end{center}
Notes: (a)~-- hydrogen column density~$\nh$ is fixed; (b)~-- the distance to the~LMC is assumed to be~$50\,\rm kpc$~\citep{Pietrzynski_etal2019}; (c)~-- the C-statistics and degrees of freedom of the best value found by the Bayesian analysis; (d)~-- eRASS1 and eRASS2 spectra were fitted simultaneously with a common WD~mass~$M$ parameter.
\end{table*}

It is important to emphasise that the obtained hydrogen column density~$\nh$ is significantly different for the~eROSITA and XMM~spectra. Fitting the XMM~spectrum provides the column density~$\nh$ is about half of that found from fitting the eROSITA spectrum. Namely, the analysis of the eRASS1~spectrum yields $\nh\,{\approx}\,5.88\times 10^{20}\,\rm cm^{-2}$, see Table~\ref{tab:fit} and Fig.~\ref{fig:corner_plot_eRASS1}, left panel. On the other side, the analysis of the XMM~spectrum yields $\nh\,{\approx}\,3.06\times 10^{20}\,\rm cm^{-2}$. The last value is too small compared to the Galactic absorption column in the direction of the source. The average Galactic absorption in this direction is $5.6_{-0.8}^{+0.7}\times 10^{20} {\rm cm^{-2}}$ based on the extinction/absorption maps presented in \citet{Doroshenko2024}\footnote{\url{http://astro.uni-tuebingen.de/nh3d/nhtool}}, who combined the results by \citet{Edenhofer_etal2024, Yao_etal2017} and \citet{Plank_etal2016}. The \ion{H}{I}~survey~\citep{HI4PI_Collaboration_2016} provides the close result $6.5_{-0.1}^{+0.1}\times 10^{20}~{\rm cm^{-2}}$. Therefore, we consider the value of~$\nh$ obtained from fitting the eROSITA~observation to be more reliable and fix it when describing the XMM~spectrum. Specifically, we fixed $\nh=5.88\times 10^{20}\,\rm cm^{-2}$. It should be also noted that we assume there is no extra~$\nh$ local to the source, so we determine only the foreground absorption component, and the T\"ubingen-Boulder ISM absorption model {\bf\tt tbabs} is used.

The corner plot and obtained parameters for the XMM~observation with the fixed~$\nh$ are presented in Fig.~\ref{fig:corner_plot_eRASS1}, right panel, and Table~\ref{tab:fit}. The comparison between the observed and model RGS~spectrum is shown in Fig.~\ref{fig:spectrum_RGS}. 

Interestingly, for both the~XMM~and eRASS1~spectra, regardless of the other parameters' values, the carbon abundance `$\chem$'~parameters are close to each other and coincide within errors, ${\approx}\,0.27$ and~${\approx}\,0.24$ respectively. They are derived quite accurately, and the corresponding distributions are narrow and do not demonstrate multi-modal behaviour. Additionally, another spectrum from the next eROSITA survey, eRASS2, is available. Fitting this spectrum with the \chem-model also yields quite a good fit with relatively small errors for~`\chem', indicating its reliable determination. This~'\chem' value is slightly lower,~${\approx}\,0.20$, than the value ${\approx}\,0.27$\,--\,$0.24$ obtained from eRASS1 and XMM~spectra.

\begin{figure*}
\centering
	\center{\includegraphics[width=1.0\linewidth]{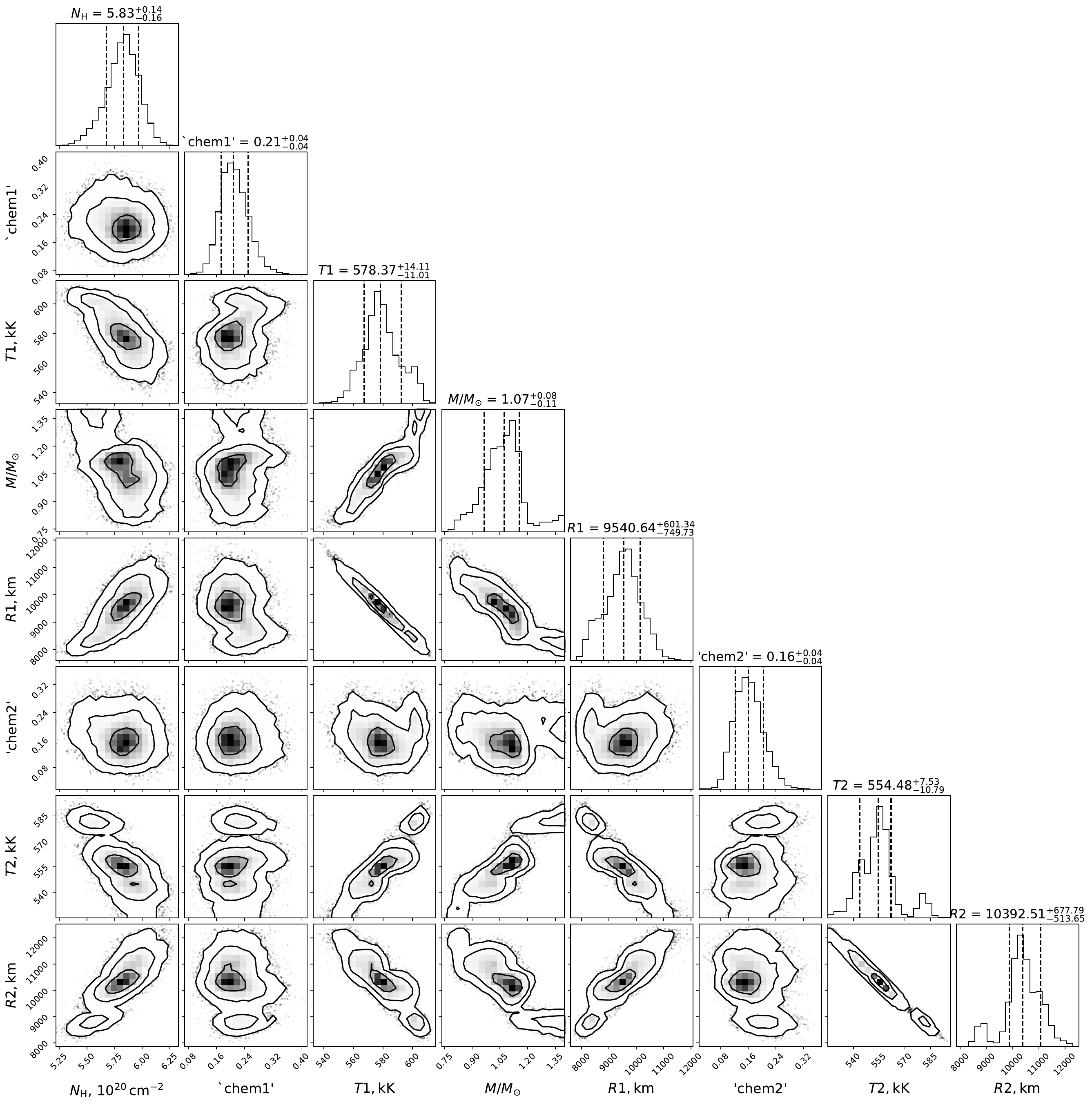}}
    \caption{Corner plot of the parameters' posterior distribution for~eRASS1 and eRASS2~spectra fitted simultaneously with a common WD mass parameter. The two-dimensional contours correspond to~$39.3\%$, $86.5\%$, and~$98.9\%$ confidence levels. The histograms show the normalised one-dimensional distributions for a given parameter derived from the posterior samples. The best-fit parameter values are presented in Table~\ref{tab:fit}.}
    \label{fig:corner_plot_eRASS12}
\end{figure*}

Given the availability of eROSITA spectra for two epochs, it is possible to trace the system parameters' evolution, particularly the carbon abundance. To achieve this, a MCMC~calculation was performed simultaneously for the~eRASS1 and eRASS2~spectra using the \chem-model, assuming that the WD mass is the same in both fits, and, therefore, it is a common free parameter for all spectra. The same uniform priors were set for the hydrogen column density, `\chem'-parameter, effective temperature, WD~mass and radius. The corresponding posterior distribution is shown in Fig.~\ref{fig:corner_plot_eRASS12}, and the parameters' values are presented in Table~\ref{tab:fit}.

The posterior distributions of the two `\chem'-parameters are quite narrow, and the corresponding errors are relatively small. Note that the hydrogen column density is a free parameter in this case, and it converged to a plausible value of~$5.83\times 10^{20}\, \rm cm^{-2}$. The inferred WD~mass,~$\approx 1.07\,M_\sun$, is also determined with quite good quality.

\begin{figure}
\centering
	\center{\includegraphics[width=1.0\linewidth]{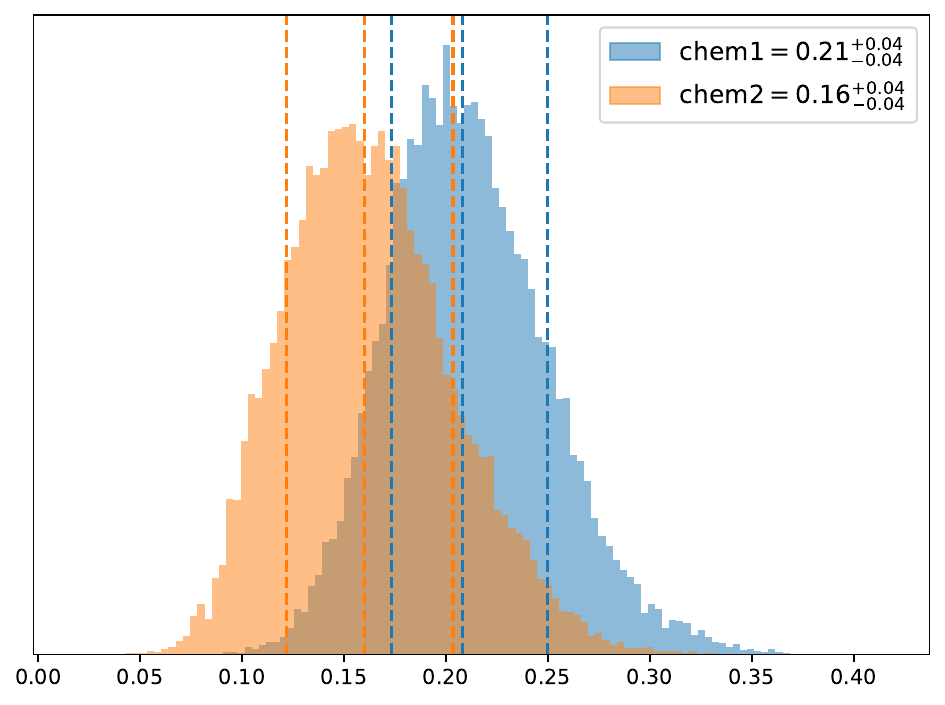}}
    \caption{One-dimensional distribution for eRASS1~(in blue) and eRASS2~(in orange) `\chem'-parameter. Spectra were fitted simultaneously using the Bayesian approach, and the WD~mass parameter was common for both spectra. Vertical lines show the~16-,~50- and~84-percentiles for `\chem'~distribution.}
    \label{fig:chem1-chem2}
\end{figure}


\begin{table}
\begin{center}
    \caption{Fit of AT~2018bej spectra with NLTE$^a$ model.}
    \label{tab:rauch_006_fit}
    \begin{tabular}{lccc}
    \hline \hline \\ [-0.5em] 
    & eRASS1 & XMM & XMM$^c$  \\ [0.3em]
    \hline \\ [-0.7em] 
    $\nh, 10^{20}\rm\,cm^{-2}$  & $4.8 \pm 0.1$ & ${\sim}\,8.54$  & $4.8$ \\ [0.3em]
    $\Teff,\,\rm kK$      &  ${\sim}\,690$ & ${\sim}\,660$ & ${\sim}\,630$ \\ [0.3em]
    $R^b,\,\rm km$        &  $3980\pm 20$ & $3970\pm 20$ & ${\sim}\,4000$ \\ [0.3em]
    $M/M_\sun$          &  $1.19\pm 0.01$ & $1.19\pm0.01$ & $1.19\pm0.01$ \\ [0.3em]
    $L, \,10^{37}\,\rm erg\,s^{-1}$   & ${\sim}\,2.5$ & ${\sim}\,2.2$ & ${\sim}\, 1.8$ \\ [0.3em]
    \multirow{2}{*}{$\texttt{cstat}\,\,{\rm (dof)}$} & $150.81$ & $7440.30$ & $15886.52$ \\ [0.3em]
    & $(128)$ & $(3325)$ & $(3326)$ \\ [0.1em]
    \hline
    \end{tabular}
\end{center}
Notes: (a)~-- \textsc{tmap} NLTE model~006 by \citet{Rauch_etal2010}, $\log g=9.0$; (b)~-- the distance to the~LMC is assumed to be~$50\,\rm kpc$~\citep{Pietrzynski_etal2019}; (c)~-- hydrogen column density $\nh$ is fixed.
\end{table}

We also compared our results with those obtained using non-LTE model spectra. A set of NLTE~atmosphere models was calculated by~\citet{Rauch_etal2010} to describe the spectrum of nova V4743~Sgr during its SSS~phase. The chemical composition of these models has been adjusted specifically for that nova. To compare our results with NLTE~models we fit our observed spectra with Rauch's model~006.
The resulting WD~radii $(2.2-3.7)\times 10^8$\,cm and masses of $0.4-1\,M_\sun$ are significantly underestimated, and the obtained radius is lower than the radius of cold WD with the similar mass. Therefore, we set a lower limit for the WD~radius using Nauenberg's $M-R$ relation \citep[see][]{Nauenberg1972}, and the obtained results are presented in Table~\ref{tab:rauch_006_fit}. The temperature reduces to~${\sim}660-690\, {\rm kK}$ while the mass increases to~${\sim}1.2\, M_\sun$. Again, we consider the~$\nh$ value found by the eRASS1~analysis to be more reliable and performed an additional fitting of the XMM~spectrum with the fixed $\nh=4.8\times 10^{20}\,\rm cm^{-2}$. The fit quality deteriorates significantly, and the mass tends to its lower limit based on the radius constraint. While the low-resolution eROSITA spectrum is more or less well described by the NLTE~model, the quality of the XMM~fit is significantly worse compared to our LTE~fit. 

It should be noted that, since the source luminosity is near the Eddington limit~($L\sim0.12-0.49\,L_{\rm Edd}$ for different LTE/NLTE fits), NLTE~effects are not negligible. However, the spectral resolution of eROSITA observations is insufficient to reveal these effects. While the XMM~resolution, in principle, allows us to distinguish NLTE~effects, they would only moderately affect the result~(e.g. by~${\sim}10\%$ in the temperature, see the end of Sect.~\ref{sect:model_atmospheres}). In addition, the available NLTE~models were computed only with fixed $\log g = 9$. Possibly, NLTE~models with smaller~$\log g$ would be able to fit the observed spectra with better accuracy. Nevertheless, \citet{Suleimanov_etal2024} compared the results of fitting the \textit{Chandra} and XMM~spectra of the classical SSS~CAL~$83$ using LTE~models and NLTE~models by~\citet{Lanz_etal2005} and concluded that both approaches yield similar results, making it possible to use LTE~models for analysing the spectra of SSSs.

\section{Discussion}
\label{sect:discussion}

\begin{figure}
\centering
	\center{\includegraphics[width=0.9\linewidth, trim={4.7mm 4.8mm 4mm 3.5mm},clip]{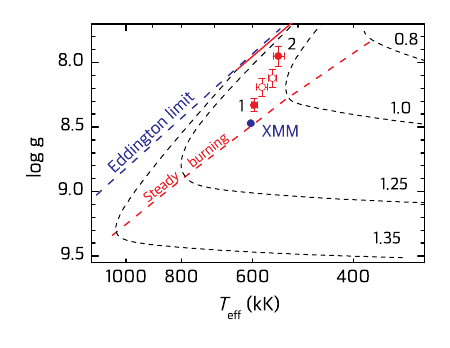}}
    \caption{Positions of the source in the $\Teff - \log g$ plane according to different eRASS observations (red circles with the number~$1$~and~$2$), and according to joint fit (red empty circles), see Table~\ref{tab:fit}. The XMM~spectrum fit with fixed~$\nh$ is shown by blue circle. Model dependencies for various WD~masses, taken from~\citep{Nomoto_etal2007}, are shown by black dashed curves. The numbers at the curves indicate WD~masses (in solar masses). The lower boundary of the stable thermonuclear burning band is shown by the dashed red line. The Eddington limit for solar H/He~abundances is shown by the blue dashed line.}
    \label{fig:glogT}
\end{figure}

Our analysis of the joint eRASS1~and eRASS2~fit allows us to investigate the evolution of the source on a half-year scale. Although the effective temperature of the hot~WD slightly decreases and the photospheric radius slightly increases, the changes in the parameter values are marginal and they coincide within errors. Formally, the carbon abundance decreases from~$0.21$~to~$0.16$ of the solar value, but the error of `\chem'~determination is comparable with the difference between these two values, as seen in Fig.~\ref{fig:chem1-chem2}. In conclusion, even the low-resolution spectral analysis shows hints of the evolution of the source parameters, including the WD's chemical composition, but observations more separated in time are necessary for a more reliable study of evolution.

It should be noted that the mass obtained from fitting different spectra varies, with a tendency to approach its upper limit. Therefore, for a more reliable mass determination, we compare the obtained results with theoretical predictions. Models of WD~envelopes with hydrogen thermonuclear burning have been computed by several groups~\citep[see e.g.][]{Wolf_etal2013, Nomoto_etal2007}. In particular, \citet{Nomoto_etal2007} calculated the $R-\Teff$ dependencies for different masses of WDs with hydrogen-rich envelope. We converted these dependencies into $\Teff - \log g$ dependencies, and put all the obtained data on the corresponding plane~(see Fig.~\ref{fig:glogT}). We included all our fits on the Figure, including the different eRASS~observations, the XMM~observation with fixed~$\nh$, and the joint eROSITA fit with a common $M$-parameter.

All the positions of the source on the plane $\Teff - \log g$ lie at the model dependencies corresponding to $1.1-1.15 \, M_\sun$, although it should be noted that the errors are large for the eRASS2~fit. This value of the WD~mass is in a good agreement with the WD~mass derived from the joint eRASS1 and eRASS2 fit. It means that the estimation of the WD mass $M=1.05-1.15\,\rm M_\sun$ could be correct. We also emphasise that all fit positions found lie above the steady-burning limit for WDs~(red dashed line in Fig.~\ref{fig:glogT}). This confirms our assumption that the source undergoes steady hydrogen burning on its surface. Moreover, the eRASS2 source position moves further away from the lower boundary of the stable burning strip, indicating an increase in the intensity of the thermonuclear burning from eRASS1 to eRASS2 observations. This fact is supported by the increased bolometric luminosity.

It should be noted that setting $\nh$ as a free parameter for the XMM~fit leads to a significant increase in gravity, causing the model mass to approach the limit mass~$1.4\, M_\sun$. As mentioned earlier, we do not consider such a fit to be satisfactory, and the corresponding point is not shown in the Fig.~\ref{fig:glogT}.

\section{Summary}
\label{sect:summary}

We performed a spectral analysis of the supersoft X-ray phase of the classical nova AT~2018bej, which was observed for the first time in X-rays by the eROSITA and \textit{XMM-Newton} telescopes. To describe the spectrum we calculated high-gravity hot LTE~model atmospheres with chemical compositions typical for nova SSS~phases. Particularly, we assumed that the number fractions of hydrogen and helium are almost equal, $\rm A_{\rm H}/A_{\rm He}=0.46/0.54$. The heavy element abundances are taken equal to the half of the solar abundances for all the elements, excluding carbon and nitrogen. We considered five different values of $\rm A_{\rm C}$ from~$0.0$ to~$0.5$ of the solar value. The nitrogen abundance is connected with~$\rm A_{\rm C}$ under the assumption that all the carbon deficiency transforms to the nitrogen excess above $0.5$~of the solar value. The grid covers $\Teff=525-700\,\rm kK$ in steps of~$25\,\rm kK$, $\dlg = \log g - \log g_{\rm Edd}=0.1,\, 0.2,\, 0.4,\, 0.6,\, 1.0,\, 1.4,$ and five values of carbon abundance (`\chem'-parameter). The code developed by \citet{Suleimanov_etal2024} was used for this aim.

Separate fits of the individual eRASS1 and eRASS2 observations reveal remarkable changes in the physical properties of the source on a half-year timescale. The effective temperature decreased, while the photospheric radius increased. Although the observed soft X-ray flux decreased, the derived bolometric luminosity slightly increased. The most 
potentially interesting change is the marginal decrease in carbon abundance between these two observations, from ${\approx}\,0.24$ to ${\approx}\,0.20$ of the solar one. The fit of the XMM/RGS spectrum yields results close to those obtained from the eRASS1~spectrum analysis if we fix the interstellar absorption.

In addition to the separate fits we performed a simultaneous fitting of the eROSITA spectra for two epochs~(eRASS1 and eRASS2) with a common WD~mass parameter. This assumption is strongly physically motivated and allows us to trace the evolution of the source more reliably. 
However, the analysis demonstrates only marginal evolution of parameters. Although the effective temperature slightly decreases, accompanied by a slight increase of the WD~radius, the bolometric luminosity is constant. A slight carbon abundance decrease was also confirmed.

We compared the WD~mass obtained from the joint eROSITA spectra fit,~$1.07^{+0.08}_{-0.11} \, M_\sun$, with theoretical predictions. The predictions were derived from the positions of the best-fit parameters of the spectra on the~$T_{\rm eff} - \log g$ plane, where model curves for different WD~masses are also displayed. The positions of the fits correspond to WD~masses of $1.1-1.15\,M_\sun$. Thus, we concluded that the mass of the WD in the investigated nova falls within the range of $1.05-1.15\,M_\sun$.

To sum up, our simplified LTE~approach provides a satisfactory description of the X-ray spectra of the supersoft X-ray stage of a classical nova. It can potentially explain the spectra of other similar objects.

\begin{acknowledgements}
    {This work was supported by the \emph{Deut\-sche For\-schungs\-ge\-mein\-schaft\/} under grants WE1312/56--1 (AT) and WE1312/59--1 (VFS). This work is partially supported by the \textsl{Bundesministerium f\"{u}r Wirtschaft und Energie} through the \textsl{Deutsches Zentrum f\"{u}r Luft- und Raumfahrt e.V. (DLR)} under the grant number FKZ 50 QR 2102 (LD). This work is based on data from eROSITA, the soft X-ray instrument aboard SRG, a joint Russian-German science mission supported by the Russian Space Agency (Roskosmos), in the interests of the Russian Academy of Sciences represented by its Space Research Institute (IKI), and the Deutsches Zentrum für Luft- und Raumfahrt (DLR). The SRG spacecraft was built by Lavochkin Association (NPOL) and its subcontractors, and is operated by NPOL with support from the Max Planck Institute for Extraterrestrial Physics (MPE). The development and construction of the eROSITA X-ray instrument was led by MPE, with contributions from the Dr. Karl Remeis Observatory Bamberg \& ECAP (FAU Erlangen-Nuernberg), the University of Hamburg Observatory, the Leibniz Institute for Astrophysics Potsdam (AIP), and the Institute for Astronomy and Astrophysics of the University of Tübingen, with the support of DLR and the Max Planck Society. The Argelander Institute for Astronomy of the University of Bonn and the Ludwig Maximilians Universit\"{a}t Munich also participated in the science preparation for eROSITA. The eROSITA data shown here were processed using the \texttt{eSASS} software system developed by the German eROSITA consortium.
    }
\end{acknowledgements}

\bibliographystyle{aa} 
\bibliography{references} 

\begin{appendix} 

\section{Additional figures}

\begin{figure}[h]
    \center{\includegraphics[width=1.0\linewidth]{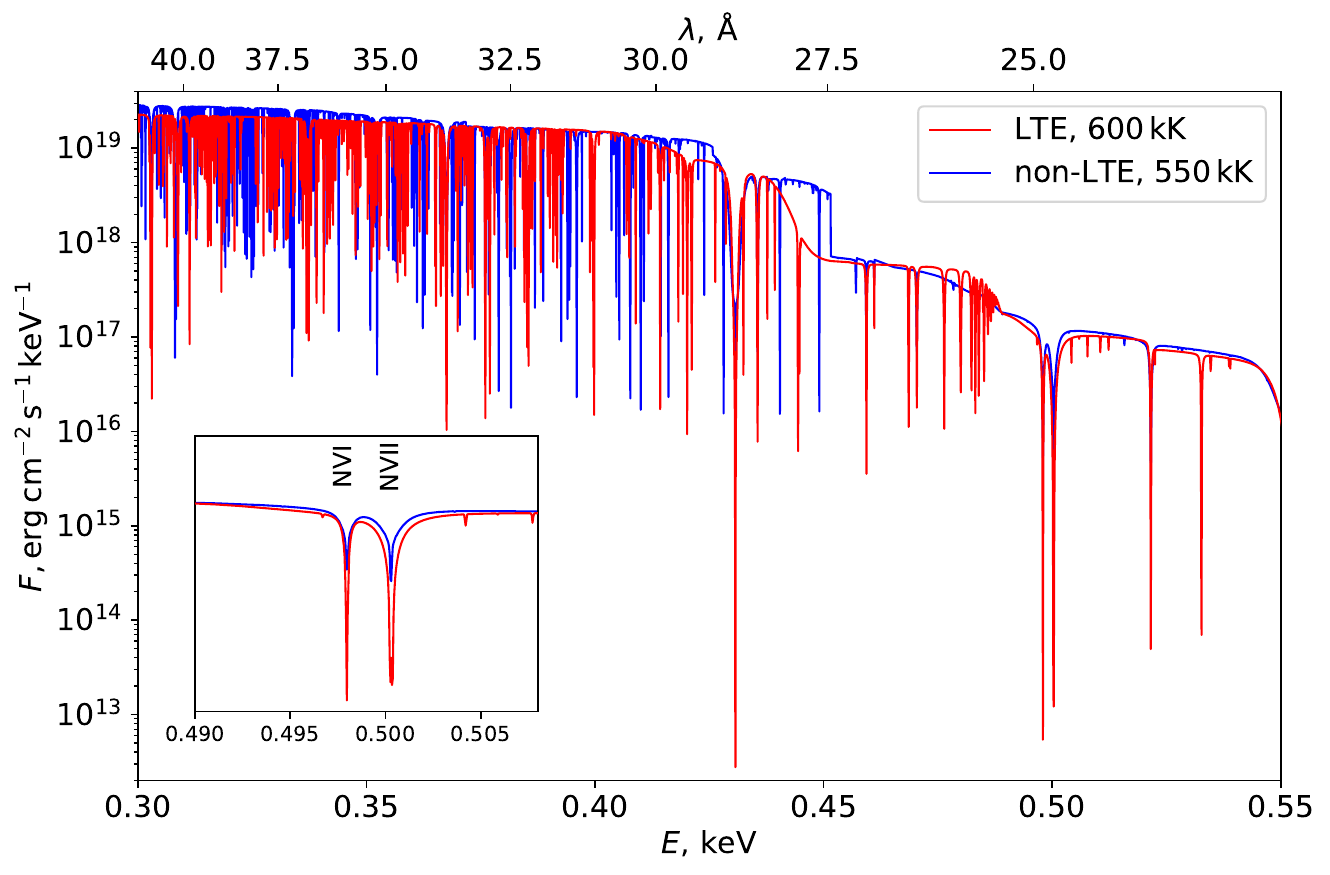}}
    \caption{Comparison of the model atmospheres spectra computed by our code~(in red) and the~\textsc{tmap} non-LTE code~\citep[][in blue]{Rauch_etal2010}.
    Shown are LTE~model with~$\Teff = 600\, {\rm kK}$ and scaled
    NLTE~model with~$\Teff=550\,\rm kK$. The gravity $\log g=9$ is fixed. The chemical composition corresponds to model~006 from Rauch's grid. The lower left panel shows in detail the comparison of absorption line profiles.}
    \label{fig:diff_rauch_550_600}
\end{figure}

\begin{figure}[h]
    \center{\includegraphics[width=1.0\linewidth]{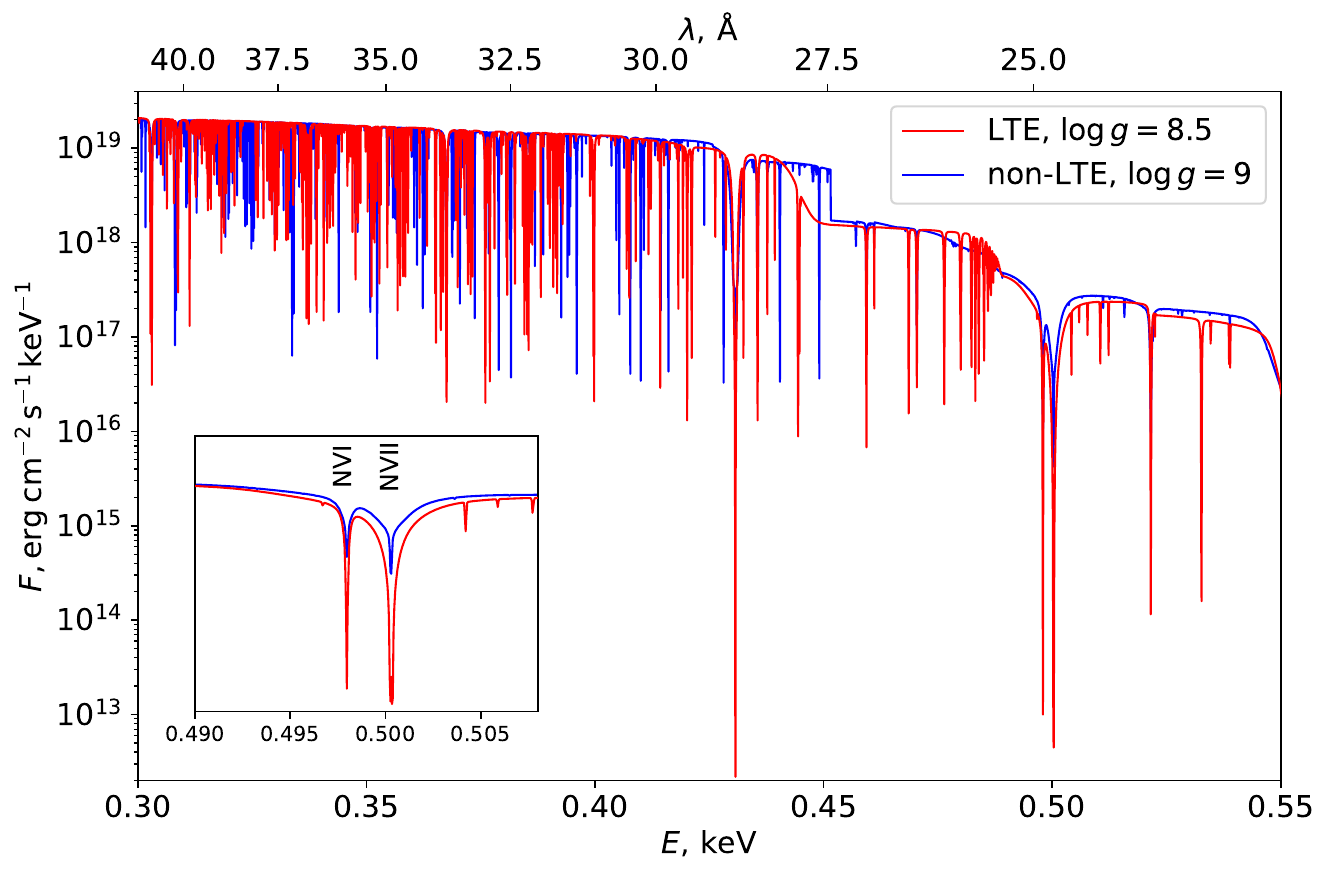}}
    \caption{Same as in Fig.~\ref{fig:diff_rauch_550_600}, but with the fixed $\Teff=600\,\rm kK$ and different gravities: $\log g=8.5$ for LTE model and $\log g = 9$ for NLTE one. 
    }
    \label{fig:diff_rauch_85_9}
\end{figure}

Here we present the results allowing us to estimate the uncertainties arising due to using LTE~model atmospheres instead of NLTE~models. The main NLTE effect important for us is over-ionisation, which leads to a decrease in the photoionisation jumps. To compensate for this effect under LTE~assumption we have to increase the plasma temperature or decrease the plasma density. We demonstrate here that a NLTE~model atmosphere with~$T_{\rm eff} = 550\,\rm kK$ reproduces the general spectral shape of the LTE~model atmosphere with $T_{\rm eff} = 600\,\rm kK$ at the same surface gravity $\log g=9.0$~(see Fig.~\ref{fig:diff_rauch_550_600}). Rauch's NLTE model~006 was used for comparison. The chemical composition of the~LTE and NLTE models is identical. The spectra are binned to a~$0.002\AA$ interval within the~$10 - 60\,\AA$ wavelength range, and the NLTE~spectrum is scaled to align directly below the LTE~one for the purpose of comparison. Is is clearly seen that the general spectrum shape, including the absorption jumps, is reproduced by the LTE~model with temperature ${\sim}\,10\%$ higher than the NLTE~one. The same effect is seen when we decrease the gravity of the LTE~model, as shown in Fig.~\ref{fig:diff_rauch_85_9}. Both spectra have the same~$\Teff=600\,\rm kK$ but the LTE~model has a lower surface gravity,~$\log g=8.5$, instead of $\log g =9$ in the NLTE model.

\end{appendix}

\end{document}